\documentclass[letterpaper,twocolumn,10pt]{article}
\usepackage{usenix,epsfig,endnotes}
\usepackage{tabularx}
\usepackage{caption}
\usepackage{subcaption}
\usepackage{multirow}
\usepackage{amsmath}
\usepackage{booktabs}
\usepackage[vlined, algo2e,ruled,linesnumbered]{algorithm2e}
\usepackage{algorithm}
\usepackage{algorithmic}
\usepackage{tikz}
\usepackage{xparse}
\usepackage{svg}

\newcommand\encircle[1]{%
  \tikz[baseline=(X.base)] 
    \node (X) [draw, shape=circle, inner sep=0] {\strut #1};}

\begin{document}

\date{}

\title{\Large \bf Aladdin: Joint Placement and Scaling for SLO-Aware LLM Serving}

\author{
{\rm Chengyi Nie}\\
Stony Brook University
\and
{\rm Rodrigo Fonseca}\\
Azure Research - Systems
\and
{\rm Zhenhua Liu}\\
Stony Brook University
}
\maketitle

\thispagestyle{empty}

\subsection*{\centering Abstract}
The demand for large language model (LLM) inference is gradually dominating the artificial intelligence workloads. Therefore, there is an urgent need for cost-efficient inference serving. Existing work focuses on single-worker optimization and lacks consideration of cluster-level management for both inference queries and computing resources. However, placing requests and managing resources without considering the query features easily causes SLO violations or resource underutilization. Providers are forced to allocate extra computing resources to guarantee user experience, leading to additional serving costs.
In this paper we introduce Aladdin, a scheduler that co-adaptively places queries and scales computing resources with SLO awareness. 
For a stream of inference queries, Aladdin first predicts minimal computing resources and the corresponding serving workers' configuration required to fulfill the SLOs for all queries. Then, it places the queries to each serving worker according to the prefill and decode latency models of batched LLM inference to maximize each worker's utilization. Results show that Aladdin reduces the serving cost of a single model by up to $71\%$ for the same SLO level compared with the baselines, which can be millions of dollars per year.

\section{Introduction}

Recently, the applications of Large Language Models (LLMs) are skyrocketing~\cite{gpts}, which greatly changes human life and work styles. The demand for LLM inference has increased significantly as more and more LLM applications become integrated into human work and life. Unlike normal Deep Neural Network inference~\cite{resnet}, which requires a small amount of GPU resources, current transformer-based LLMs consist of billions of parameters. This makes LLM inference highly dependent on expensive GPU resources, specifically on GPU memory and computing power. The current state of LLM applications has led to a shortage of GPU resources in both public and private clouds~\cite{gpu_shortage}. With this common sense, efficiently managing and scaling the GPUs for LLM inference becomes a vital problem.

To improve the efficiency of LLM inference, Previous work~\cite{responselength} considers scheduling the requests with similar predicted output lengths to one batch for efficient batch inference, recent work~\cite{orca,vllm,sarathi} focus on efficient dynamic batching for LLM inference to address the problem that requests in one batch have different output lengths. FlexGen~\cite{flexgen} improves the LLM inference by aggregating CPU and GPU resources. Splitwise, etc~\cite{splitwise, distserve, interference} separate the prompt processing stage and token generation stage into different instances to optimize the throughput and goodput. They adapt naive algorithms like Join the Shortest Queue (JSQ) to place requests for workers. Previous work~\cite{interference} adapts a power-of-two algorithm for the request placement according to the predicted output length. However, they all focus on improving the LLM inference throughput, and some improve the Service Level Objectives (SLO) attainment as a side effect. Previous work~\cite{clockwork, mark, infaas,clipper} investigated the SLO-aware DNN serving. However, the workload for those work is highly predictable. The early work on workload placement and resource management~\cite{tetris, Morpheus} has a deep dive into cluster-level management. However, they focus on the traditional workload, which is distinct from the characteristics of LLM inference jobs. To the best of our knowledge, there is no previous work that guarantees the SLOs for all LLM queries as well as improving the request placement and worker scaling for optimized inference service cost.

In continuous batching inference, the KV cache usage of each request increases while decoding and becomes zero when the token generation is finished. The peak of KV cache usage of each request is right before the end of decoding. If all requests in a batch finish decoding simultaneously, the KV cache can easily overflow. Good request placement has to prevent this situation. When we implement continuous batching, a feature of the decoding phase becomes apparent: the decoding latency increases as more tokens are generated and stored in the KV cache, even with the same batch sizes. Simply constraining the batch size for the decoding phase can also result in violations of the decoding SLO. However, current solutions lack awareness of those features.

The management and scaling of workers also significantly affect the cost of LLM inference. A worker serves as the smallest unit for inference. The demand for LLM inference varies throughout the day. For example, in the daytime, the demand is higher, necessitating more workers to meet the inference SLOs. Conversely, the demand decreases at nighttime, allowing for a reduction in the number of workers to save on inference costs.
Regarding the cluster configuration, we aim to address the following question: What is the minimum number of GPUs required to serve LLM queries while meeting all SLOs? This involves considering two decision variables: the number of GPUs per worker and the total number of workers. The current LLM inference system~\cite{splitwise} configures one inference worker with all GPUs on a machine. However, this static configuration is suboptimal for most models. DistServe~\cite{distserve} considered the Goodput of each GPU that optimized each worker's configuration from the computing latency perspective. However, it does not consider the key-value (KV) cache constraints of worker configuration or harness the features of arrival queries for workload allocation.

\begin{figure}[h]
    \centering
    \includegraphics[width=0.33\textwidth]{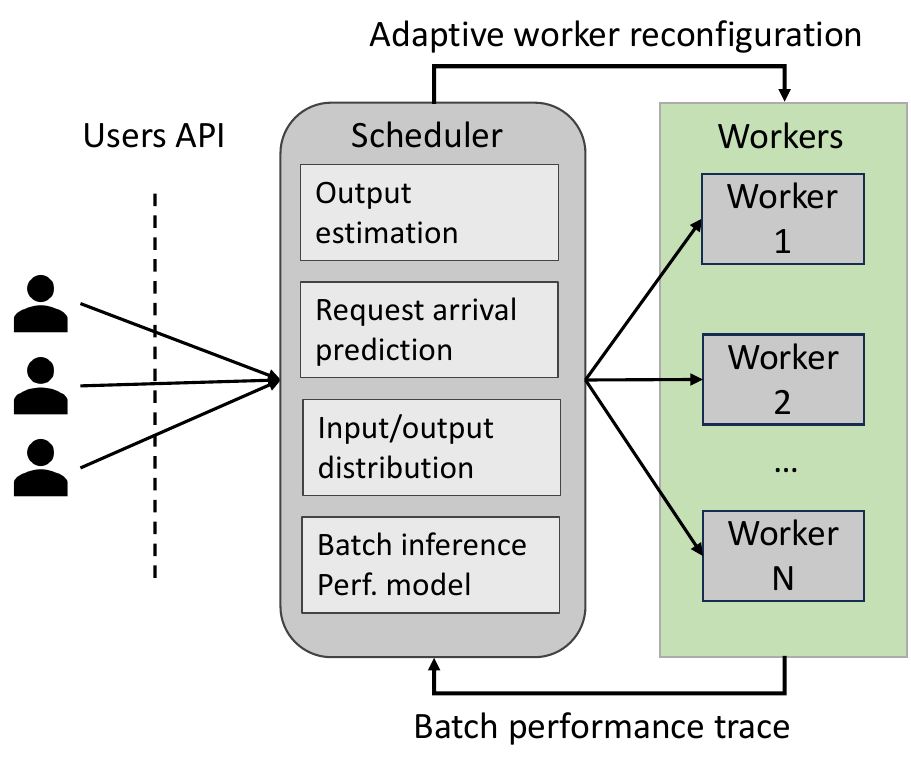}
    \caption{The overall architecture of co-adaptive scheduling}
    \label{fig:intro}
\end{figure}

Based on the insights and limitations of the literature, we propose Aladdin, a co-adaptive scheduler for request placement and resource scaling. As shown in Figure~\ref{fig:intro}, when LLM inference requests arrive, Aladdin first predicts minimal computing resources by learning the optimal configuration of serving workers based on the historical input-output length distributions and the request arriving rate. Secondly, Based on the requests' input and predicted output length, as well as the learned batching performance models, we formulate the request placement to an online multi-dimensional bin packing problem. Lastly, We monitor the ongoing requests of each worker and adjust the placement of new arrivals to reduce the impact of output length prediction errors. 
Aladdin supports the default setting vLLM~\cite{vllm} that does the prefill and decode in the same worker, as well as the decoupled prefill and decode setting like~\cite{splitwise, distserve, interference}.

Overall, the main contributions of our paper are:
\begin{itemize}
    \item We conduct an empirical study of the dynamic batching performance of prefill-decoding LLM inference and deduce the accurate performance prediction model of LLM serving.
\end{itemize}
\begin{itemize}
    \item We design a near-optimal online algorithm and a novel scheduler, Aladdin, to co-adaptively place the queries and manage computing resources to fulfill all requests' SLOs using minimal GPUs.
\end{itemize}
\begin{itemize}
    \item We conducted a comprehensive evaluation of Aladdin, including the validation of our LLM inference performance models on the A100 and V100 testbeds to establish its generality. We evaluated Aladdin's end-to-end performance with the real-world workload, which arrived as a stream on GPU servers. Additionally, we conducted a large-scale simulation for the high-demand LLM serving scenario. 
\end{itemize}

\section{Background and Motivation}
\subsection{Batch Processing of LLM Requests}
\label{background:linear models}
The demand for large language model (LLM) serving has experienced exponential growth, making the efficient serving of LLM requests a critical challenge. LLM serving places significant demands on GPU computing power and memory, which can be prohibitively expensive. Previous work, such as Orca~\cite{orca} and vLLM~\cite{vllm}, have introduced dynamic continuous batching techniques for transformer-based generative models to optimize GPU utilization.

LLM generates responses iteratively, producing one token at a time and using it as input for the next iteration. Each request generates one token after every iteration in a batch of LLM requests. Importantly, these requests may have varying output lengths, necessitating different numbers of iterations to complete. Traditional request-level batching methods pose a disadvantage. Requests within the same batch must wait until all requests are finished before results are returned. In contrast, continuous batching employs iteration-level scheduling, submitting an iteration calculation to the execution engine with each token generation. This approach prevents early-finish requests from waiting for the completion of other requests, improving GPU utilization.

\subsection{LLM Inference SLOs }
\label{sec:slo}
In contrast to other DNN inference workloads~\cite{clockwork} that have well-defined latency targets, LLM inference is a two-stage iterative process. The first stage involves the generation of the initial token, which processes all prefilled tokens, while the second stage is the decode stage, where tokens are generated iteratively one by one. LLM inference latency depends on the output length. Although the time for generating the first token increases with the number of prefilled tokens~\cite{sarathi}, it remains predictable based on the length of the prefilled tokens. Additionally, the first token generation is a single-round inference process without iteration, so we have set a predetermined response deadline for time to the first token (TTFT). 

For the decoding process, previous work~\cite{splitwise} adopts the time between tokens (TBT) metric, constraining the latency between every token smaller than the target. However, the TBT metric is an over-strict metric with less flexibility, and it does not directly affect the user's quality of experience. We introduce the quality of experience SLO using the average token generation time (ATGT) metric $\textit{ATGT}=\frac{t_{decode}}{l_{out}-1}$, where $t_{decode}$ is the decode time of a request and $l_{out}-1$ is the output length of the decode phase. This metric reflects the average time spent generating each token during the decode stage. For example, the average reading speed for individuals is approximately four words per second~\cite{readspeed}. To ensure the delivery of quality service, the average token generation time for each request must not exceed 0.2 seconds.

\subsection{Output Length Prediction}
The input and output lengths of requests have a huge impact on the decision of the inference requests and worker configuration. However, when we make the request placement decisions, we only have the information for the input length of each request. There are some techniques to predict the output length of each request. Previous work~\cite{responselength, interference, sche_out_pred} proposed the response length perception that harnesses the output length prediction before the execution ability of LLMs. They use historical data to fine-tune the LLM. However, there are drawbacks to this methodology. Firstly, the overhead of using a LLM to predict the output length is non-negligible because the output length prediction process is another inference. Although previous work~\cite{interference} uses a smaller model to predict the output length for a larger LLM, the prediction overhead is still significant. And the prediction of response length perception is out of control. From our experiment result, the response length predicted by the fine-tuned models is biased. 

\begin{figure}
    \centering
    \includegraphics[width=0.9\linewidth]{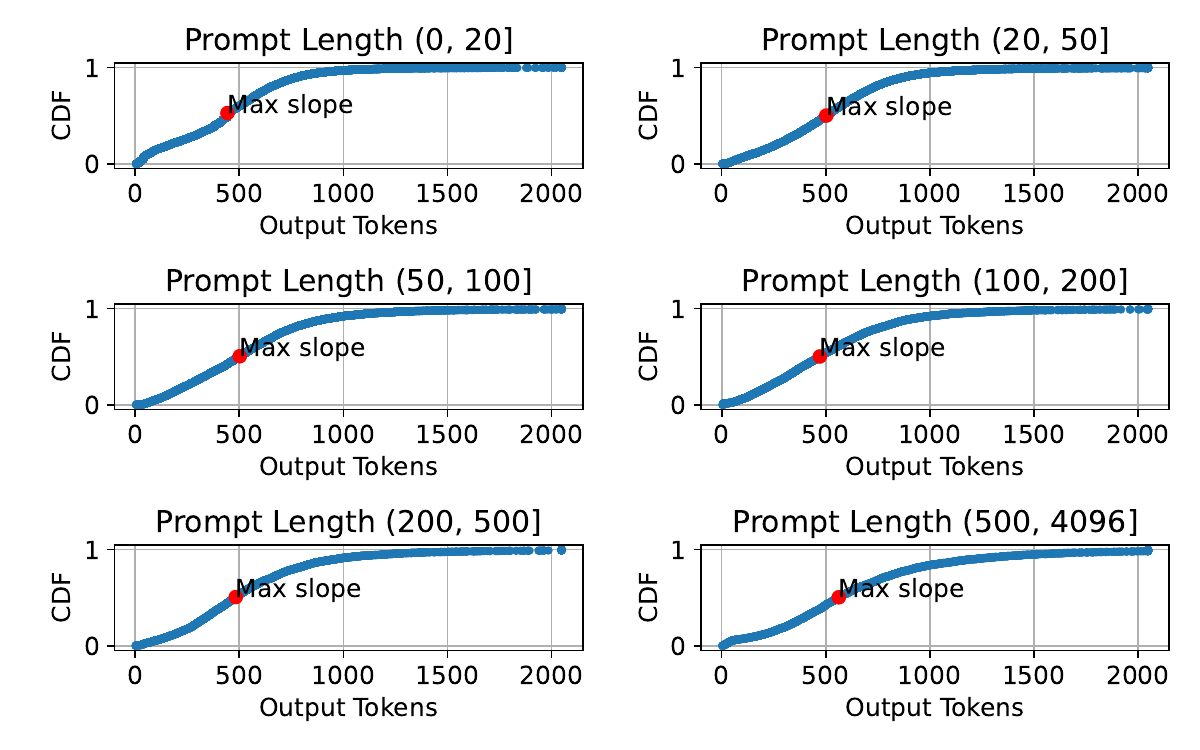}
    \caption{CDF of output length for different prompt Lengths from ShareGPT and llama2-13b-chat-hf generated output.}
    \label{fig:inout_cdf}
\end{figure}

Figure~\ref{fig:inout_cdf} presents the CDF of output length given the corresponding prompt length in different ranges. Although the output length prediction error is inevitable in our request placement, the prediction without bias can partially cancel the prediction error when we put requests in a batch.  Hence, we use the estimated output length of each input length in the historical data as the predicted output length. This is the most naive output length predictor. Although the prediction error may be high, this prediction method has a low overhead and is non-biased. 
In Section~\ref{sec:error balancing}, we address the prediction error by designing a novel re-balancing algorithm. Note that the output length prediction is not the main contribution of this paper. If there are accurate, non-biased, and low overhead output length predictors in the future, the performance of Aladdin could be further improved.

\subsection{Challenges and Opportunities}
There are challenges to improving the request placement and worker scaling.\\
\textbf{Challenge 1: Heterogeneous phases of LLM inference.} The transformer-based LLM inference consists of prefilling and decoding stages. The prefill stage is the first iteration of an inference request that processes all prompt tokens; it has more computing demand than the decoding process. The decoding process is a memory-intensive stage compared with the prefill stage because of the KV cache. These distinct features result in different performance models of prefilling and decoding processes for each request. Given the requests with various input and output lengths, accurately predicting the iteration time of batched prefill and decode is challenging. \\
\textbf{Challenge 2: Worker performance prediction.} The inference workload varies over time with high uncertainty. Meanwhile, worker configuration and the number of workers directly affect the cost of inference. Considering the request arrival pattern, we must take into account the worker's computing latency, KV cache capacity, and communication overhead. The search space for configurations is too large to be explored by a naive enumeration approach. Accurately predicting optimal configurations poses significant challenges. \\
\textbf{Challenge 3: Handle the error of output length prediction.} The output length prediction error is inevitable. Therefore, reducing the impact of prediction errors on output length is crucial for enhancing performance when assigning tasks to workers. Systems need to effectively react when the prediction error is detected.

We tackle the request placement problem by transforming it into a multi-dimensional bin packing problem. As the LLM inference process is predictable, we develop a dynamic batching inference performance model. We consider the arrival pattern of queries with the output length prediction error awareness. Since the first token response time, average token generation time, and the KV cache demand are predictable, they facilitate the design of the scheduling algorithm. 

With a thorough analysis of the computing and communication overhead of tensor parallelism and batching execution, we demonstrate the predictability of the inference throughput and the latency at the iteration level. In Aladdin's design, we predict the most resource-efficient worker configuration according to the performance of GPUs with their interconnection, LLM model size, and SLOs. With this optimal worker configuration, we can reach the highest SLO attainment rate with the same GPU resource. Furthermore, to achieve additional cost savings, we dynamically adjust the number of workers based on trends in arrival rates and query features.

\section{Continuous Batching Performance Modeling}
\label{sec:modeling}
\subsection{KV Cache Usage} 
In LLM inference, The transformer uses the given prompt (context) as the initial input and generates additional tokens one by one. During the inference process, the transformer performs self-attention, which requires the key-value (KV) vectors for each token (prompt and generated tokens) in the current sequence. These vectors are stored in the GPU as two matrices (key matrix and value matrix) during inference, often called the KV cache.
At the beginning of an inference, the KV cache stores the key and value matrices of the prompt tokens. During response generation, the KV vectors associated with that token are appended to the KV cache matrices with each token generated. This dynamic expansion leads to a linear relationship between the KV cache's usage and the current sequence size. This linear relationship signifies that the KV cache's memory footprint increases proportionally with the sequence length. 
So the KV cache usage of a request
\begin{equation}
\label{eq:kv cache}
    kv = h(l_{in}+l_{out}) +j,
\end{equation}
where $h$ and $j$ are learnable coefficients, and $r$ is the output tokens generated so far.

\subsection{Iteration Time} 
Iteration-level batching poses unique challenges. Not all requests can be batched together at any iteration due to varying input shapes. Orca~\cite{orca} addresses this by proposing selective batching.
However, operators like Attention require inputs with identical shapes, leading to separate calculations using cuBLAS~\cite{cublas} routines for batch matrix multiplication. The separate multiplications for each request result in a linear scaling of iteration time to the batch size. In default settings like vLLM~\cite{vllm} or split-phase inference, one batch can only contain prefill or decode. Since the query in the attention mechanism of the prefill process is a matrix that includes all input tokens, the query of the decode process is a vector of the last generated token. The iteration latency model of the prefill and decode batch is different.\\
\textbf{Prefill iteration time.} Since prompt processing is a computing-bottleneck process, a single request with a reasonable input length can effectively saturate the worker's computing power, which means the batching effect has limited improvement to the throughput in the prefill process. Our preliminary results indicate that the iteration time of the prefill batch is not affected by the batch size and is linear with the total input length of all batched requests. The iteration time:
\begin{equation}
\label{eq:prefill iter time}
    t_{pre} = k_1\sum l_{in}+ c_1,
\end{equation}
where the $\sum l_in$ is the total input length of all requests in the prefill batch, $k_1$ and $c_1$ are the learnable coefficients. \\
\textbf{Decode iteration time.} However, the token generation process has low compute utilization since each query only generates one token in an iteration. 
With a fixed batch size, the iteration time linearly increases as the average context length (the input length of the request and the tokens generated so far) increases. Similarly, with the same average context length, the iteration time increases linearly with the batch size. According to the experiment, the iteration time with a batch size of one (i.e., single request inference without batching) remains nearly constant. 
With this information, when we haven't reached the KV cache limit, the iteration time $t_{d}$ is:
\begin{equation}
\label{eq:iter time}
    t_{d} = (k_2l_{ave}+c_2)b + c_3, \ b>1,
\end{equation}
where $b$ is the batch size, $l_{ave}$ is the average context length among all requests. $k$ and $c$ are learnable coefficients.
In the scheduling algorithm design, given the ATGT SLO $T_{dec}$, the total input length is limited by a function of batch size $b$:
\begin{equation}
\label{eq:input limit}
    l_d \leq \frac{1}{k_2}\left(-c_2b + T_{dec} - c_3\right), \ b>1.
\end{equation}
Note that all coefficients in Eq.~\ref{eq:input limit} are positive according to the batch inference scaling. And $T_{dec}$ must be greater than $c_3$ because the decoding latency SLO we choose must be greater than the individual request decoding latency without batching. From Eq.~\ref{eq:input limit}, we deduce that with a larger batch size, the maximum total input length limit of all requests within the batch decreases.

\section{Co-Adaptive Scheduling}
When requests arrive at the scheduler, our task is to determine how to use the minimum number of GPUs to serve both newly submitted and ongoing requests while ensuring compliance with the SLO requirements. This overarching objective can be deconstructed into several critical components:
\begin{itemize}
    \item We need to determine the minimal GPU number required to serve the queries that fulfill the SLO requirements.
\end{itemize}
\begin{itemize}
    \item Find the most efficient configuration of these GPUs, such as the number of workers and the number of GPUs configured with each worker.
\end{itemize}
\begin{itemize}
    \item Decide how to place the requests to each worker in a manner that optimizes the utilization of each worker.
\end{itemize}
It's important to note that these three components are interconnected. When one decision is made, the other two are simultaneously determined. For example, when we establish the total number of GPUs, this decision implicitly dictates the optimized placement of GPUs and models on each worker, as well as the optimization of request assignments to each worker. Conversely, if we can devise a more effective strategy for worker configuration or request assignment that enhances resource utilization, we can reduce the total resource requirements for a more cost-efficient service. Firstly, Let's look into the optimal single-worker configuration because the optimal configuration for each worker is orthogonal to the request scheduling and worker number determination.
\subsection{Worker Configuration}
\label{sec:worker configuration}
Model parallelism is a widely used technique for LLM training and inference. There are two types of model parallelism: tensor parallelism, which splits the tensors across all GPUs, and pipeline parallelism, which splits the layers across all GPUs. In general, people use tensor parallelism inside a node where GPUs are connected by high-bandwidth networks like NVlink, and pipeline parallelism is used when there is slower cross-node communication. However, the performance modeling is challenging for pipeline parallelism because of bubbling. In this paper, we consider the tensor parallelism distributed inference. The optimal worker configuration is achieved when we achieve the optimal per-GPU throughput. Therefore, the throughput with the given number of GPUs is optimized. With the different ranks of tensor parallelism, the computing, communication, and KV cache capacity all impact the throughput. In the default vLLM~\cite{vllm} setting, the prefill and decode processes are served with the same worker. The decode process dominates the inference process because tokens are generated in the decode process one by one while the prefill process only has one iteration. We have to predict the parallelism strategy with the most per-GPU throughput for decode phase.

In tensor parallelism, each GPU first computes the split tensor locally, then combines the tensor across all GPUs using All-reduce. The split tensor size is inverse to the GPU number, so the computing time is an inverse function of the number of GPUs:
\begin{equation}
\label{eq:TP compute}
    t_{compute} = \frac{k_4}{N_g}+ c_4,
\end{equation}
where $N_g$ is the number of GPUs per worker, $k_4$ and $c_4$ are learnable parameters. Tensor parallelism adopts All-reduce for the inter-GPU communication. The communication overhead for All-reduce, relative to the number of GPUs, is $(N_g-1)/N_g$. When the number of GPUs is large, the communication overhead of All-reduce is nearly constant. However, for modern GPU servers like DGX A100 and H100, the number of GPUs on each server is less than or equal to $8$. So the difference in the communication overhead is non-negligible. Since the communication between GPUs is intra-node communication with a high-speed network. The communication straggler is not significant. Therefore we use the communication overhead to predict the communication delay scaling to the GPU number. The KV cache capacity can be calculated by the sum of the GPUs' memory on each worker minus the model size, $M = N_gm_{gpu}-m_{model}$. The throughput can be limited by KV cache or the SLO constraint. When KV cache is the bottleneck, the maximum throughput is achieved when the KV cache is full. When iteration SLO is the bottleneck, the maximum throughput is achieved when the decode iteration time is equal to ATGT SLO latency limit. 
The maximum per-GPU throughput of tensor parallelism rank $N$ is:
\begin{equation}
\label{eq:max throughput}
    T_{max}=min\left\{\frac{M}{N_gm_r(t_{compute} + t_{comm})}, \frac{B}{N_gT_{decode}}\right\},
\end{equation}
where $m_r$ is the average per request KV cache demand learned from the historical data, and $t_{compute} + t_{comm}$ is the iteration time given the batch size $\frac{M}{m_r}$ with $N_g$ GPU per worker. $T_{decode}$ is the ATGT SLO, and $B$ is the batch size corresponding to the SLO. The optimal worker configuration has $N_g^{opt}$ GPUs that maximize $T_{max}$. Note that the optimal worker configuration remains unaffected by the request arrival rate but is influenced by factors such as the model's size, context length, and the computing and memory capacity of the GPU. Since we consider homogeneous GPUs in this paper, when scheduling requests and workers to adapt to varying workloads, the configuration of each worker remains unchanged.

\subsection{Request Placement Policies}

\begin{figure}
    \centering
    \includegraphics[width=0.9\linewidth]{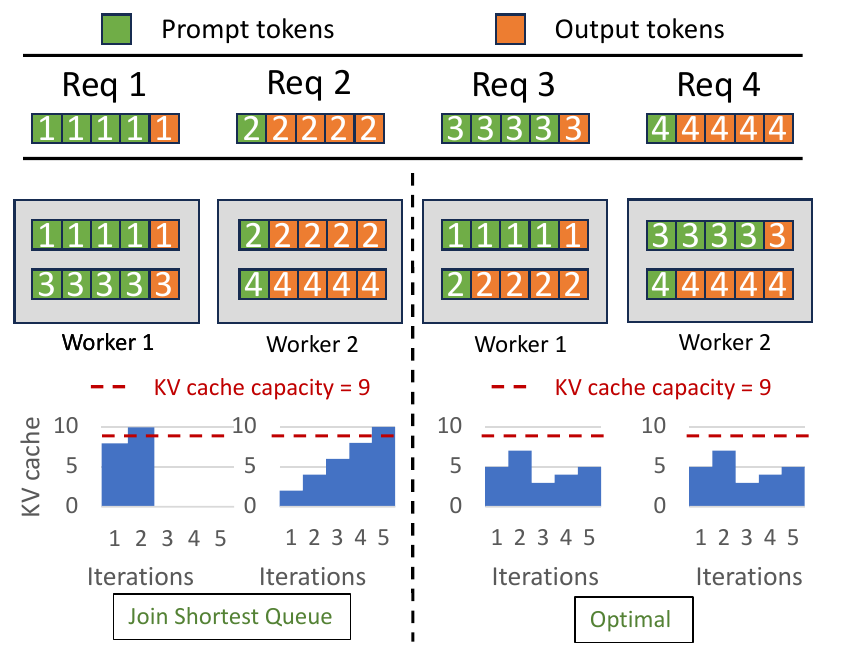}
    \caption{An example illustrates the sub-optimal of JSQ for request placement. }
    \label{fig:example}
\end{figure}

 We optimized the worker configuration to achieve maximum per-GPU throughput, and our next objective is to minimize the number of workers required for LLM service. The placement of queries to workers significantly affects efficiency of resource utilization. Figure~\ref{fig:example} illustrates the suboptimal of naive JSQ and reveals the optimal request placement strategy. In this example, requests need to be placed to two workers with KV cache capacity of $9$. Note that in this example, the requests arrive in sequence from $1$ to $4$ but are submitted to workers at the same time. If we adopt JSQ, two long prompt requests will be placed to the same worker, while two long output requests will be placed to another worker. Suppose a token requires $1$ KV cache capacity. The max KV cache demand for both workers is $10$ when requests finish generation, which exceeds the KV cache capacity of $9$. Therefore, we need to move requests to the waiting queue until there is available KV cache. However, with the optimal request placement, a long prompt request and a long output request are placed in one worker. The max KV cache demand for each worker is $7$. To aid in this decision-making process, we leverage the parameters notated in Table~\ref{tab:input&output} and the following information:
\begin{itemize}
    \item Learnable prefill time to total input tokens Eq.~\ref{eq:prefill iter time}, input tokens limit to batch size when constraining the decode iteration time~Eq.\ref{eq:input limit} and learnable KV cache usage to token count~Eq.\ref{eq:kv cache} functions for each group.
\end{itemize}
\begin{itemize}
    \item The current KV cache usage $m=\sum kv$ and total KV cache $M$ for each worker.
\end{itemize}
\begin{itemize}
    \item For each newly added request, we utilize the known input prefill length $l^{in}_j$ and predicted output length $l^{pred}_j$. For ongoing requests, we take into account the current length generated $l^{out}_j$.
\end{itemize}

\begin{table}[h]
\caption{The inputs to Aladdin and decisions Aladdin makes}
  \footnotesize
  \begin{tabular}{cll}
    \toprule
    Inputs & Notation & Definition   \\ 
 
    \midrule
  & $kv(t)$ & The KV cache usage to tokens function  \\
  &  $l_{d}(b)$ & The input length limit to batch sizes  \\
  &  $t_{p}(l)$ & The prefill iteration time function  \\
  & $m_i$ & The KV cache usage of Worker $i$, $i \in W$ \\
  &  $M$ & The KV cache capacity of each worker\\
  &  $l^{in}_j$ & The input length of a request  \\
  &  $l^{pred}_{j}$ & The predicted output length of a request  \\
  &  $l^{real}_{j}$ & The real output length of a request  \\
  &  $l^{out}_j$ & The output tokens a request generated so far   \\
  &  $t^{dec}_j$ & The time spent for decoding phase so far   \\
  &  $T_{pre}$ & The SLO of prefill latency   \\
  \toprule
  Outputs & Notation & Definition   \\
  \midrule
  &  $W$ & The total worker number  \\
  &  $x_{ij}$ & binary variable for request $j$  \\
  &  $y_{i}$ & binary variable for Worker $i$  \\
  \bottomrule
  \end{tabular}
  \label{tab:input&output}
\end{table}

The request scheduling with the constraints can be seen as a multi-dimensional bin packing problem. We formulate it as a mixed integer programming (MIP) that schedules the new-arrived requests between the scheduling heartbeat with different input/output lengths $l_{in}$ and $l_{out}^{pre}$, and we want to minimize worker number $W$. 

Let $x_{ij}$ be a binary variable that equals $1$ if request $j$ is scheduled to Worker $i$, and $0$ otherwise. Let $y_i$ be a binary variable that equals $1$ if Worker $i$ is used, and $0$ otherwise. Assume $I$ is the initial worker number larger than the optimal $W$. When there are ongoing requests, for an ongoing request $j$, to prevent the unnecessary migration between workers, the $x_{ij}$ is kept the same as the current decoding worker. We also need to guarantee that the new request's prompt processing time won't cause the token generation time SLO violation of the ongoing requests. The MIP problem can be formulated as follows:
\begin{align*}
    \min\quad &\sum_{i=1}^{I}y_i \\
    \text{s.t.}\quad &\sum_{i=1}^{I}x_{ij}=1, j=1, 2, \dots, J, &\encircle{a} \\
    &\sum_{j=1}^{J} x_{ij}(l_{j}^{in}+\gamma l_j^{out}) \leq \theta l^d\left(\sum_{j=1}^{J} x_{ij}\right) , i=1, 2, \dots, I, &\encircle{b}\\
    & t_{p} \left(\sum_{j=1}^{J_{new}}  x_{ij}l_{j}^{in}\right) \leq T_{pre}, i=1, 2, \dots, I, &\encircle{c}\\
    & t_{p} \left(\sum_{j=1}^{J_{new}} x_{ij}l_{j}^{in}\right) \leq \theta min(T_{dec}l_{ij}^{out}-t_{ij}^{dec}), i=1, \dots, I, &\encircle{d}\\
    &\left[\sum_{j=1}^{J}{\bf w}_j x_{ij}\right]_k \leq M, k=1, 2, \dots, K, i=1, 2, \dots, I, &\encircle{e}\\
    &x_{ij} \leq y_i, i=1, 2, \dots, I, j=1, 2, \dots, J, &\encircle{f}\\
    &x_{ij} \in \{0,1\}, i=1, 2, \dots, I, j=1, 2, \dots, J, &\encircle{g}\\
    &y_i \in \{0, 1\}, i=1, 2, \dots I. &\encircle{h}
\end{align*}
The constraints are: \encircle{a} Each request must be scheduled to one worker. \encircle{b} According to Eq.~\ref{eq:iter time}, the iteration time is determined by both batch size and the total context length. Eq.~\ref{eq:input limit} shows the maximum total context length of all requests in one batch given the batch sizes. This constraint ensures the ATGT SLO for the decode process. Since the iteration time increases as more tokens are generated during decoding, the coefficient $\gamma$ can be considered as a "strictness knob" that tunes the scheduling bound, $0 \leq \gamma \leq 1$. When $\gamma=0$, only the first iteration can meet the ATGT SLO. When $\gamma=1$, the last token generation time can meet the ATGT SLO. We normally set $\gamma=0.5$ to increase the worker utilization while guaranteeing the SLOs. \encircle{c} According to Eq.~\ref{eq:prefill iter time}, the sum of all new requests' input is limited by the TTFT SLO. \encircle{d} Since the prefill of new requests preempts the decode for ongoing requests, the prefill time of new requests can not exceed the time that ongoing requests have saved compared with the ATGT limit. Reflecting on the limitation of the sum of new requests' input length. \encircle{e} The total KV cache demand of all the requests scheduled to each worker cannot exceed the KV cache capacity $M$. $K$ is the sequence length limit of the serving model. $\bf{w}$ is the vector with length $K$ that shows a request's KV cache footprint. For example, for request $j$, 
\begin{equation*}
    {\bf{w}}= \begin{bmatrix}kv(l_j^{in})&kv(l_j^{in}+1)&\cdots&kv(l_j^{in}+l^{pred}_j)&0&\cdots&0\end{bmatrix},
\end{equation*}
where each element in the vector presents the KV cache demand of an iteration. The KV cache demand for the first iteration includes the KV cache for input tokens. The KV cache demand increases in the following iterations while output tokens are generated. The KV cache demand becomes zero when the request $j$ finishes. This constraint guarantees that for all scheduled iterations, the KV cache demand will not exceed the KV cache capacity of the worker.  \encircle{f} If a worker is used, it should have at least one request scheduled. Otherwise, we don't need this worker. \encircle{g}\encircle{h} All variables are binary. Unused boxes will have $y_i=0$ and will not be counted in the objective function. $0<\theta<1$ in \encircle{b}\encircle{d} is another hyperparameter that adapts to the prediction error of output length. For example, when $\theta$ is small, the constraints are tighter, so requests are less likely to violate the SLOs. However, the drawback is that we need more workers for the serving.

\begin{algorithm2e}[t]
  \caption{Request scheduling heuristic}
  \label{alg:heuristic}

  {\bfseries Input:} $l^{in}$, $l^{pred}$ of the new request $j$. $l_{in}$, $l_{pred}$, $l_{out}$ of all ongoing requests. KV cache capacity $M$ for each worker. Worker number $W$. Performance models $kv(t)$, $t_{iter}(b,l)$, $t_{pre}(l)$.\\
  {\bfseries Output:} Worker $i$ where job $j$ be scheduled, $x_{ij}=1$.\\
   {\bfseries Initial:} \textit{workerfound} $\leftarrow False$\\

   Sort all bins on \textit{capacity\_norm} from large to small. \\
   \For{sorted bins $i=1,2,\dots, I$}{
   \textbf{initial} $x_{ij} \leftarrow 0, i=1,2,\dots, I$\\
   $x_{ij}$=1\\
   \If{\encircle{b} \textbf{and} \encircle{c} \textbf{and} \encircle{d} \textbf{and} \encircle{e} \textbf{for} $i$}{
   \textit{workerfound} $\leftarrow True$\\
   \textbf{return} $x_{ij}$
   }
   }
   \If{\textit{workerfound} $=False$}{
   Open a new bin $(I+1)$ and add job $j$.\\
   \textit{workerfound} $\leftarrow True$\\
   \textbf{return} $x_{(I+1)j}=1$
   }
   
\end{algorithm2e}

\noindent \textbf{Scheduling heuristic.} 
The multi-dimensional bin-packing problem is NP-hard, so an efficient heuristic is needed to approach optimal scheduling. Given that requests arrive in an online pattern, we employ the best-fit algorithm for online bin packing~\cite{ffd}. It schedules each arrived request to the worker with the maximum load and can guarantee the satisfaction of all SLO constraints. Intuitively, this heuristic increases the utilization of each worker compared to other scheduling algorithms, such as joining the shortest queue, thereby reducing the number of workers needed.

In the multi-dimensional bin packing problem, determining the metric for each worker's load is non-trivial. Using the batch size of each worker as the metric for its load is sub-optimal because the input and output lengths of requests significantly influence each worker's load. We propose \textit{capacity\_norm}, which is the L2 norm of batch size $B$ and weighted context length $\sum (l_{in}+\gamma l_{out})$ of all ongoing requests to rank all workers. The heuristic algorithm for scheduling an arriving request is described in Algorithm~\ref{alg:heuristic}.

\begin{algorithm2e}[t]
  \caption{Re-balancing with prediction error}
  \label{alg:pre_error}

  {\bfseries Input:} $x_{ij}, l_j^{pred}, l_j^{out}, l_j^{real}$ of $J_{old}$ ongoing requests. $x_{ij}, l_j^{in}, l_j^{pred}$ of $J_{new}$ new requests. \\
  {\bfseries Output:} Updated $x_{ij}$ of new requests.\\
  {\bfseries Initial:} $l_i^e=b_i^e=0, i=1,2,\dots, I.$ \\
  \For{worker $i=1,2,\dots, I$}{
      \For{ongoing job $ j=1,2,\dots, J_i$ on worker $i$}{
          \textcolor{red}{/*Check if under estimate output length*/}\\
          \If{$l_j^{out}>l_j^{pred}$}{
               $l_i^e \leftarrow l_i^e+l_j'^{pred}$\\
               $b_i^e \leftarrow b_i^e + 1$\\
            }
            \textcolor{red}{/*Check if over estimate output length*/}\\
           \If{$l_j^{real}<l_j^{pred}$}{
                 $l_i^e \leftarrow l_i^e+l_j^{real}-l_j^{pred}$\\
               $b_i^e \leftarrow b_i^e - 1$ \\
                }
          }
      }

    Calculate the equivalent error function $\alpha_i l_i^e+ \beta_i b_i^e + c_1=0$ of worker $i, i = 1,2,\dots, I.$ according to Eq.~\ref{eq:input limit}.\\
    \textcolor{red}{/*Fix error by adjusting the new requests placement*/}\\
    \If{new request $j$ from worker $x$ to worker $y$}{
        $b^e_x \leftarrow b^e_x -1$ \\
        $b^e_y \leftarrow b^e_y +1$ \\
        $l^e_x \leftarrow l^e_x-l_j^{pred} $ \\
        $l^e_y \leftarrow l^e_y+l_j^{pred} $ \\
        
    }
    \textcolor{red}{/*Minimize the sum of the shortest distance between each worker's error function and the origin.*/}\\
    $min(\sum \frac{|c_i|}{\sqrt{\alpha_i^2+\beta_i^2}}), i=1,2,\dots, I.$\\
    {\bfseries Return} $x_{ij}, j=1,2, \dots, J_{new}$

\end{algorithm2e}

\subsection{Addressing Prediction Errors}
\label{sec:error balancing}
The output length cannot be accurately predicted before execution. If we overestimate the output length, worker utilization will be reduced. Conversely, there will be SLO violations. 
When an ongoing request in a batch finishes earlier than predicted, we mark this worker as overestimated. If an ongoing request's output length is underestimated, i.e., the request hasn't finished with the predicted tokens, we mark this worker as underestimated and predict the output length again. Before the execution of the new requests, we re-schedule new requests that have been scheduled to the over-utilized workers to the under-utilized workers. We use $l^e$ and $b^e$ as the metrics to indicate the estimation error of each worker, where $l^e$ is the accumulated error of output length for outstanding requests, and $b^e$ is the error of batch size for each worker. 
If Request $j$ is finished before the estimated iteration, which means we overestimate the output length, we can calculate the output length over-estimate error $l_j^{real}-l_j^{pred}$. If we underestimate the output length of Request $j$, we predict the output length $l_j'^{pred}$ again using conditional average output length when $l_j^{real}>l_j^{pred}$ with the same input length $l_j^{in}$. 
In the request scheduling, we use $l^e$ and $b^e$ as the indicators to balance the workload between workers and reduce the effect of output length prediction error. The calculation for $l^e$, $b^e$, and the re-balancing algorithm are described in Algorithm~\ref{alg:pre_error}.

\section{System Design}
Benefiting from the predictable nature of individual and batch LLM inference, we attempt to reveal the best way to serve requests that arrive as a stream from resource management and request placement perspectives. In this section, we describe the system design of Aladdin for two variances settings: default continuous batching and split-phase inference. The default continuous batching will process the input tokens and generate output tokens in the same worker, represented by vLLM~\cite{vllm}. The split-phase inference refers to the inference setting that splits the prompt processing and token generation into different working instances, and each instance only processes prompt or generates output. This setting is represented by Splitwise~\cite{splitwise} and DistServe~\cite{distserve}.

\begin{figure}[t]
    \centering
    \includegraphics[width=0.7\linewidth]{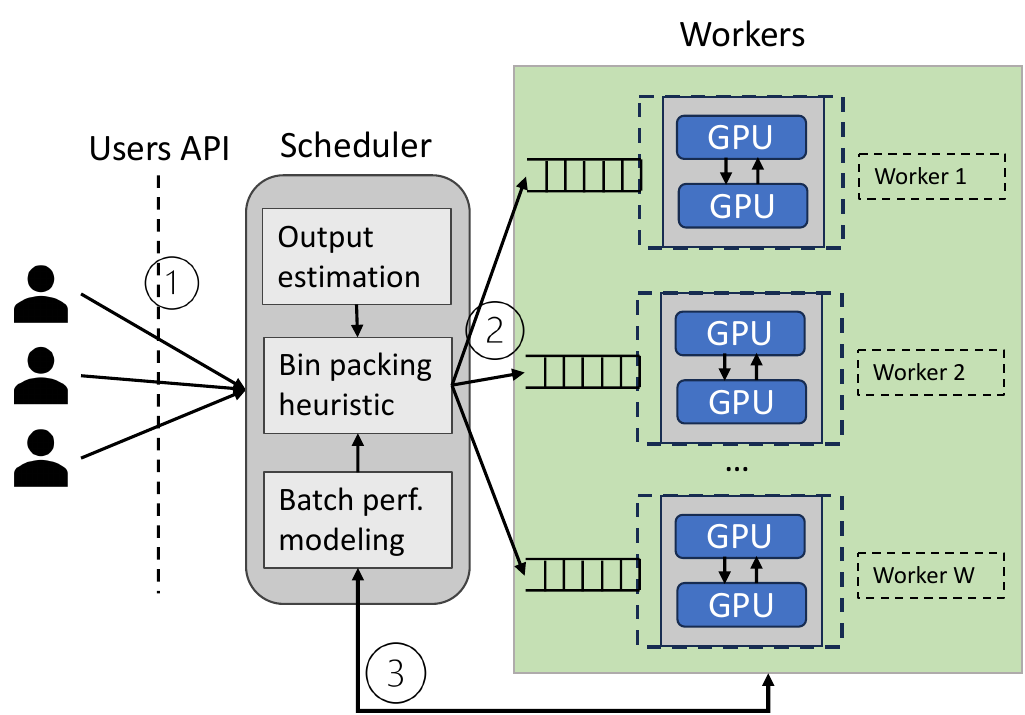}
    \caption{Workflow of Aladdin with default continuous batching}
    \label{fig:workflow_default}
\end{figure}

\begin{figure}[t]
    \centering
    \includegraphics[width=1\linewidth]{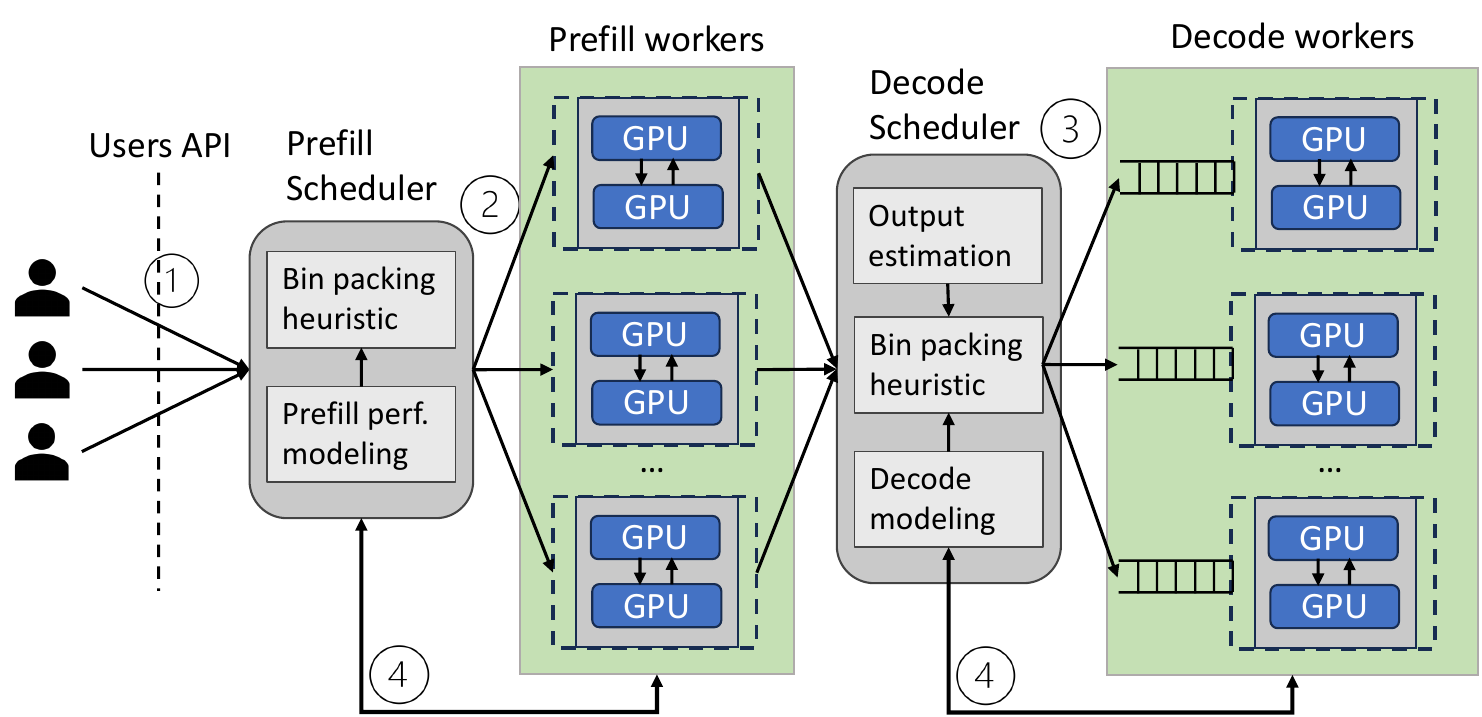}
    \caption{Workflow of Aladdin with split-phase inference.}
    \label{fig:workflow_split}
\end{figure}

\subsection{System Workflow.} 
\textbf{Default continuous batching.} The Figure~\ref{fig:workflow_default} illustrates the workflow of continuous batching inference scheduling. Firstly, users submit their LLM inference requests via the API as the first step \encircle{1}. The request scheduler uses the bin packing heuristic to schedule the new requests according to their input length and the predicted output length \encircle{2}. Lastly, the request scheduler continuously update the performance model according to the worker's execution traces \encircle{3}.\\
\textbf{Split-phase inference.} Figure~\ref{fig:workflow_split} illustrates the workflow of split-phase inference. Users submit requests through API \encircle{1}. We schedule the prefill of new requests based on their input lengths. Since the prefill only involves one iteration, there is no queue for the prefill workers \encircle{2}. Next, the decoding scheduler places the requests from prefill workers to decoding workers based on the predicted output length and a learned performance model \encircle{3}. Finally, the prefill and decode schedulers continuously update the performance model according to the workers' execution traces \encircle{4}.

\subsection{Adapt to Changing Demand}
\label{sec: adapt_workload}
In every cluster heartbeat, we can reconfigure the cluster using change point detection. In LLM inference, although users submit different queries and receive different answers, the input and output lengths of LLM inference requests for the same model exhibit a strong pattern. From the SharGPT dataset~\cite{vicuna2023}, we found that the input lengths of user queries follow a fixed distribution, and the output lengths of the same LLM also follow a learnable distribution. According to our experiment using Algorithm~\ref{alg:heuristic}, when the arrival rate $r_a$ is larger than a lower bound $R$, the total number of required workers $N_w$ is linear with the request arrival rate $r_a$.
\begin{equation}
    \label{eq:total_worker_pre}
    N_w=\lceil k_5r_a+c_5 \rceil,\ r_a>R
\end{equation}
where $k_5$ and $c_5$ are learnable coefficients associated with the historical demand, and we round the number of workers to the smallest integer larger than the function of $r_a$. The reason $R$ exists is that when the arrival rate is lower, there are fewer requests arriving in the same heartbeat, which cannot represent the real distributions of the input and output length. The standard error of the mean $SEM=\frac{\sigma}{\sqrt{n}}$ is the metric for the difference between the sampled requests' input and output lengths and the total requests, where $\sigma$ is the standard deviation of all requests' input and output length and $n$ is the number of requests we place during a heartbeat. The smaller $n$ is, the more error appears in the prediction of $N_w$.

With this model, we can predict the total number of workers required before placing all requests to each worker. However, the scheduling time requirement of inference serving is in milliseconds. In a high-demand situation, the scheduling overhead is too large to schedule the requests in the target iteration for the centralized scheduler. We design a distributed scheduler for the high-demand scenario that harnesses the pattern of input and output length of requests in Appendix~\ref{sec:dist sched}. 

Note that in this paper, we focus on predicting the minimal GPU required for the varying arrival rate without considering the cold start problem and the switching cost. Since the optimization of cluster scheduling is orthogonal to the worker number prediction problem, we defer it to future work.

\begin{figure}

    \centering
    \begin{subfigure}[t]{0.24\textwidth}
        \centering
        \includegraphics[width=\linewidth]{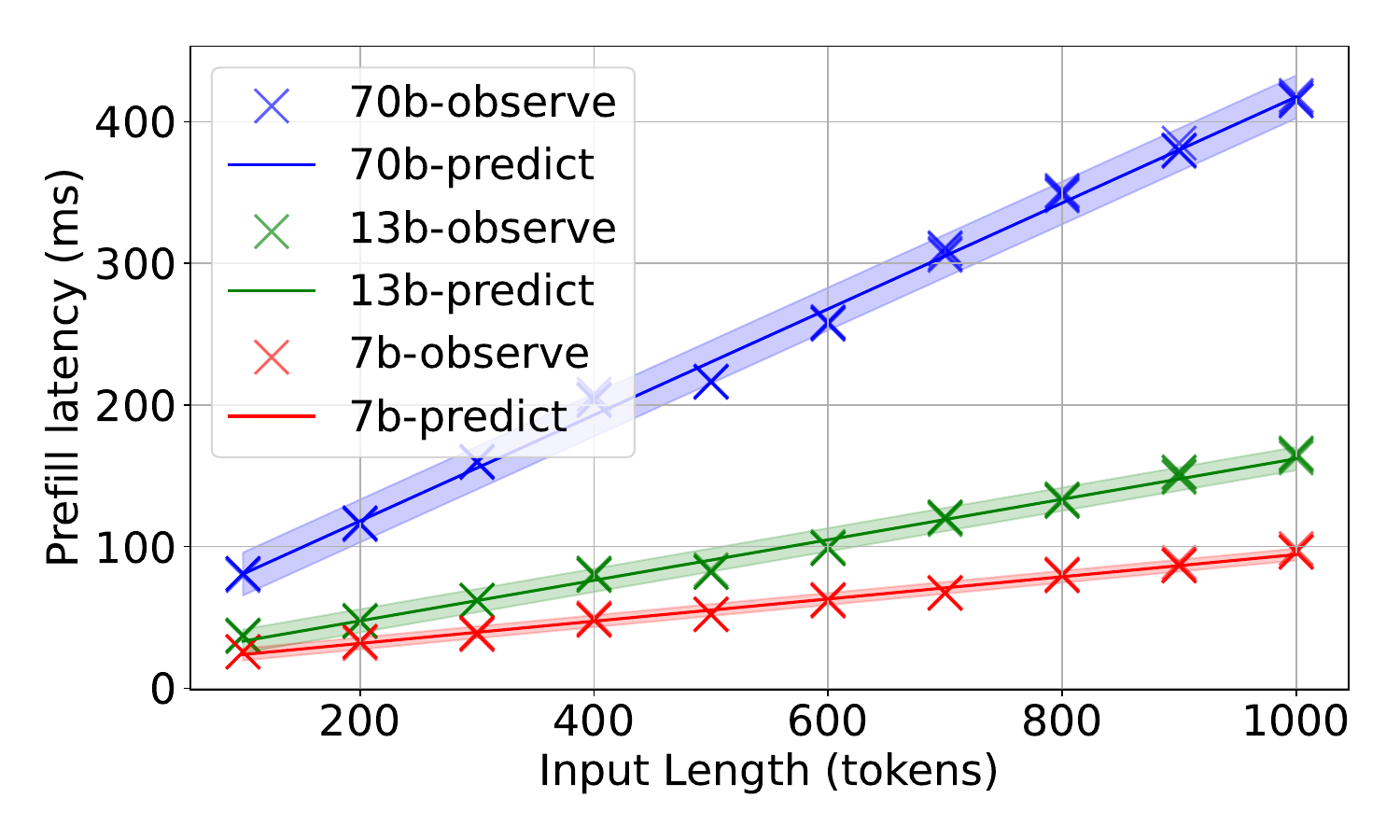}  
        \caption{A100 testbed}
        \label{fig:prefill_latency_a100}
    \end{subfigure}%
    \begin{subfigure}[t]{0.24\textwidth}
        \centering
        \includegraphics[width=\linewidth]{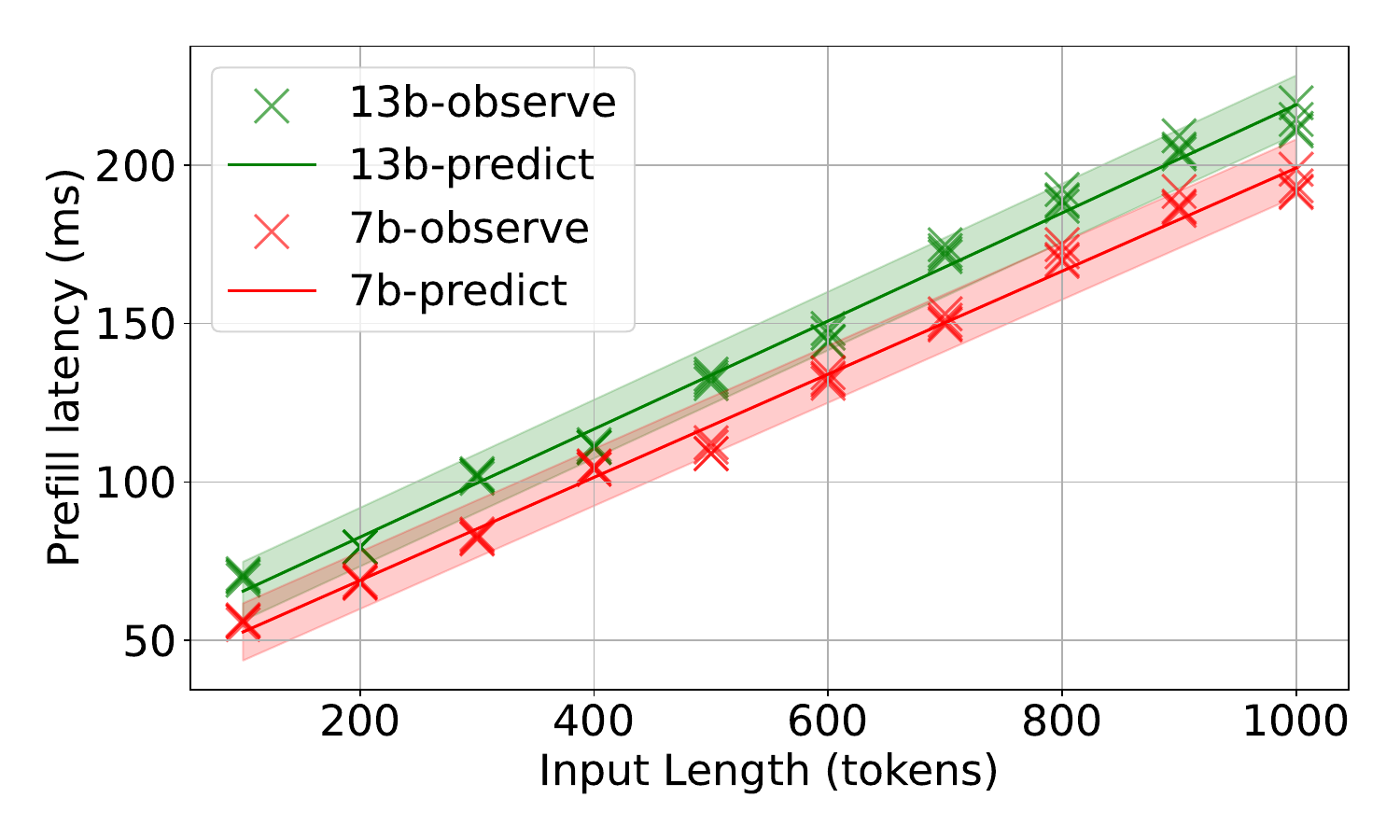}  
        \caption{V100 testbed}
        \label{fig:prefill_latency_v100}
    \end{subfigure}
    \caption{Prefill latency}
\end{figure}

\begin{figure}

    \centering
    \begin{subfigure}[t]{0.24\textwidth}
        \centering
        \includegraphics[width=\linewidth]{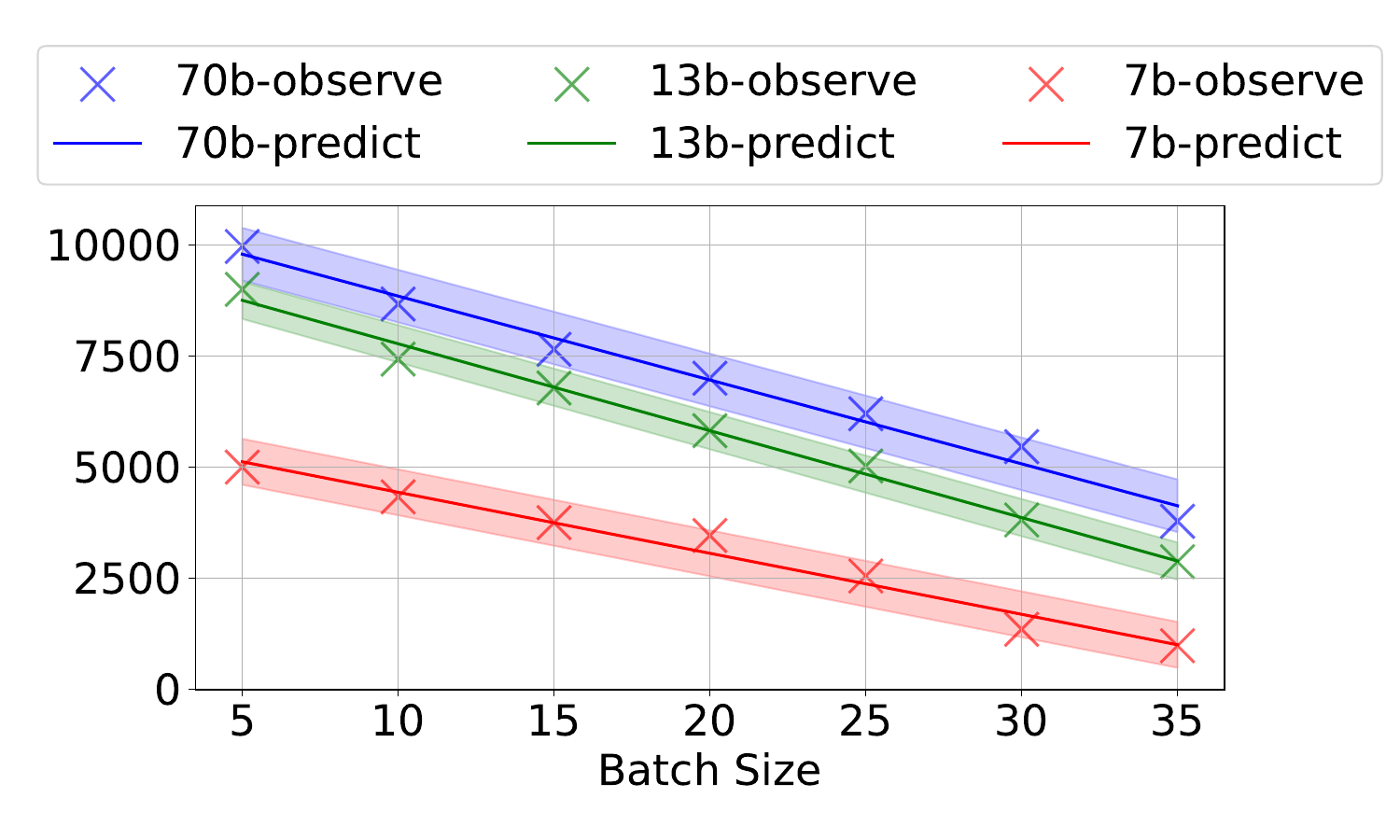}  
        \caption{A100 testbed}
        \label{fig:decode limit a100}
    \end{subfigure}%
    \begin{subfigure}[t]{0.24\textwidth}
        \centering
        \includegraphics[width=\linewidth]{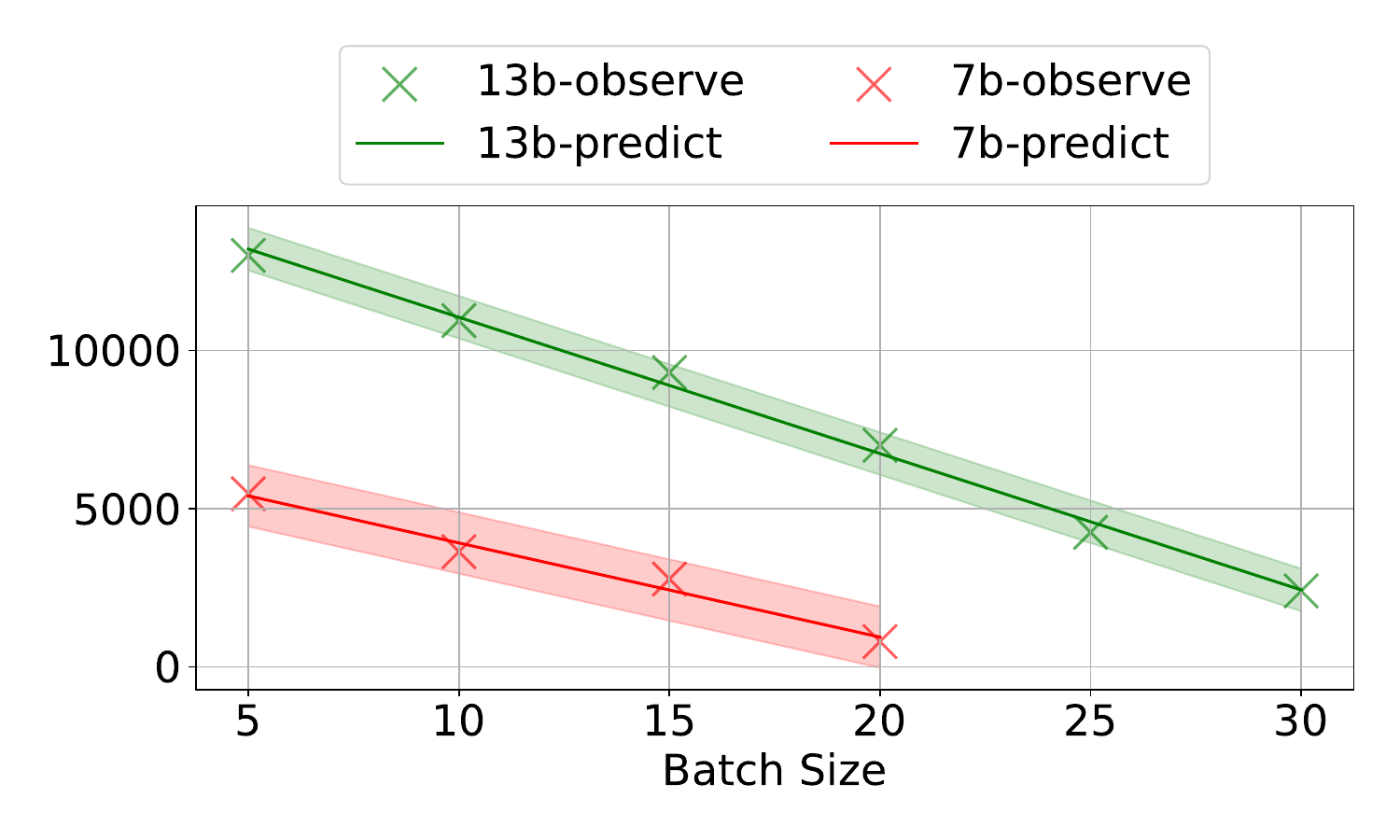}  
        \caption{V100 testbed}
        \label{fig:decode limit v100}
    \end{subfigure}
    \caption{Decode context length limitation}
\end{figure}

\subsection{Implementation}
Aladdin is specifically designed for single-model serving, eliminating any model cold start problem for each worker. We adopt vLLM~\cite{vllm} for dynamic batch inference to optimize the KV cache usage of each worker and make the KV cache usage more predictable. Aladdin's request scheduler is a scheduling layer on top of the vLLM inference engine. Users submit their requests to the Aladdin frontend through the API interface. Aladdin routes and schedules the requests to different workers through each server's API interface. Note that Aladdin is a non-blocking system; once a request is scheduled to a worker, it will start inference in the next iteration. Aladdin doesn't support request migration, which means once a request has been sent to a worker, we won't migrate it to another worker with the same duty. 

\section{Evaluation}
\label{section:eval}

\begin{figure}
    \centering
    \includegraphics[width=0.7\linewidth]{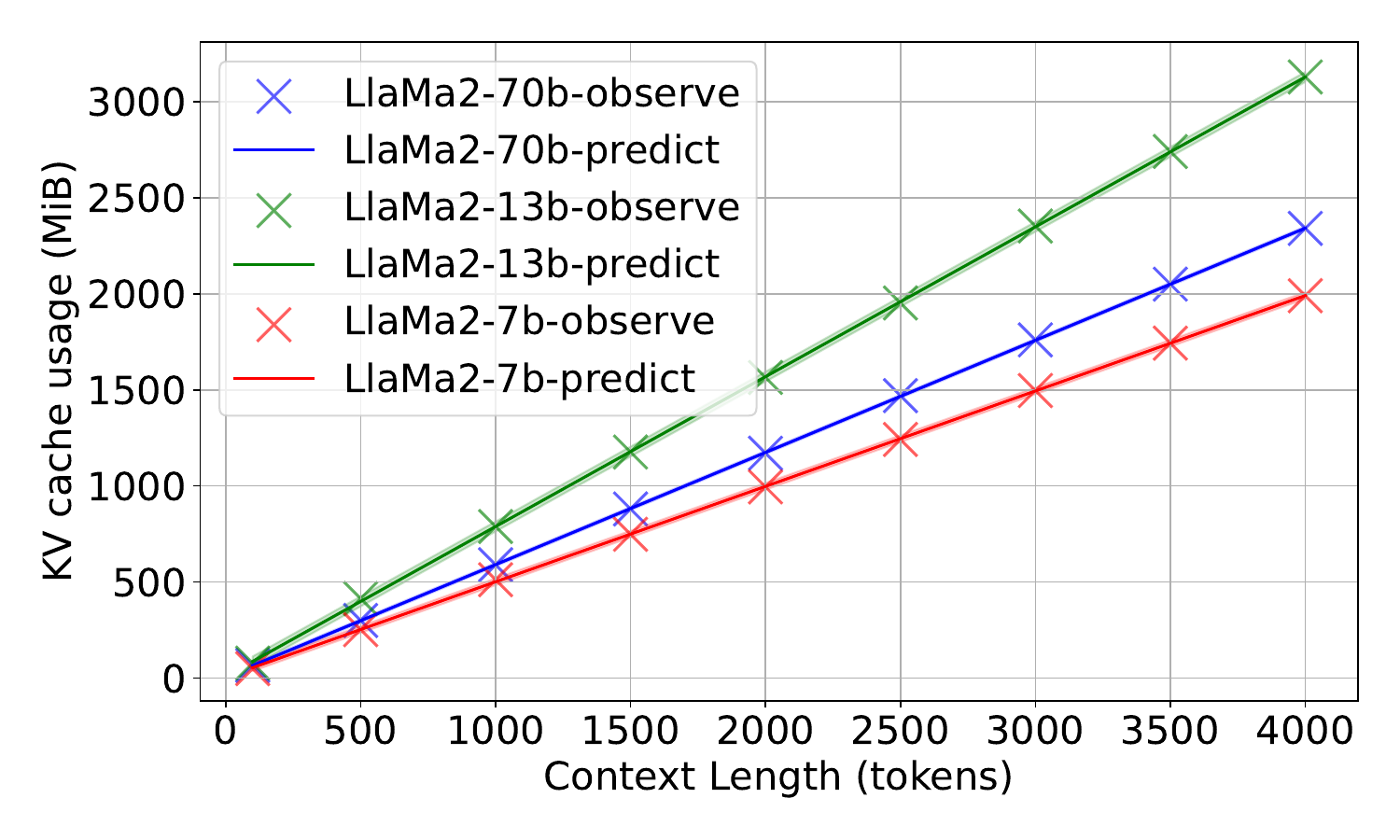}
    \caption{KV cache prediction and the observation.}
    \label{fig:kv_usage}
\end{figure}

For the evaluation of Aladdin, our first step is to validate the accuracy of our performance modeling for continuous batching inference in Section~\ref{sec:model validate}. Next, we examine the performance improvement achieved with Aladdin with different scenarios in Section~\ref{sec:endtoend} and Section~\ref{sec:simlulation}. We also provide the overhead analysis of Aladdin in Section~\ref{sec:sche_overhead}. The primary information of our evaluation is as follows:
\begin{itemize}
    \item Aladdin accurately predicts performance metrics with the maximum error less than $10\%$.

    \item Aladdin reduces the GPU number required by up to $71\%$ and $60\%$ compared with vanilla vLLM~\cite{vllm}, and split-phase inference engines~\cite{distserve, splitwise}'s decode instances for the same workload. 

    \item Although single-worker optimization techniques like chunked prefill~\cite{sarathi} and split-phase inference~\cite{splitwise, distserve} reduce the cost for inference, the cost reduced by Aladdin is orthogonal to those techniques. Aladdin can be combined with those single-worker optimization techniques to improve the performance further.
\end{itemize}

\begin{figure}
    \centering
    \begin{subfigure}[b]{0.4\textwidth}
        \centering
        \includegraphics[width=\textwidth]{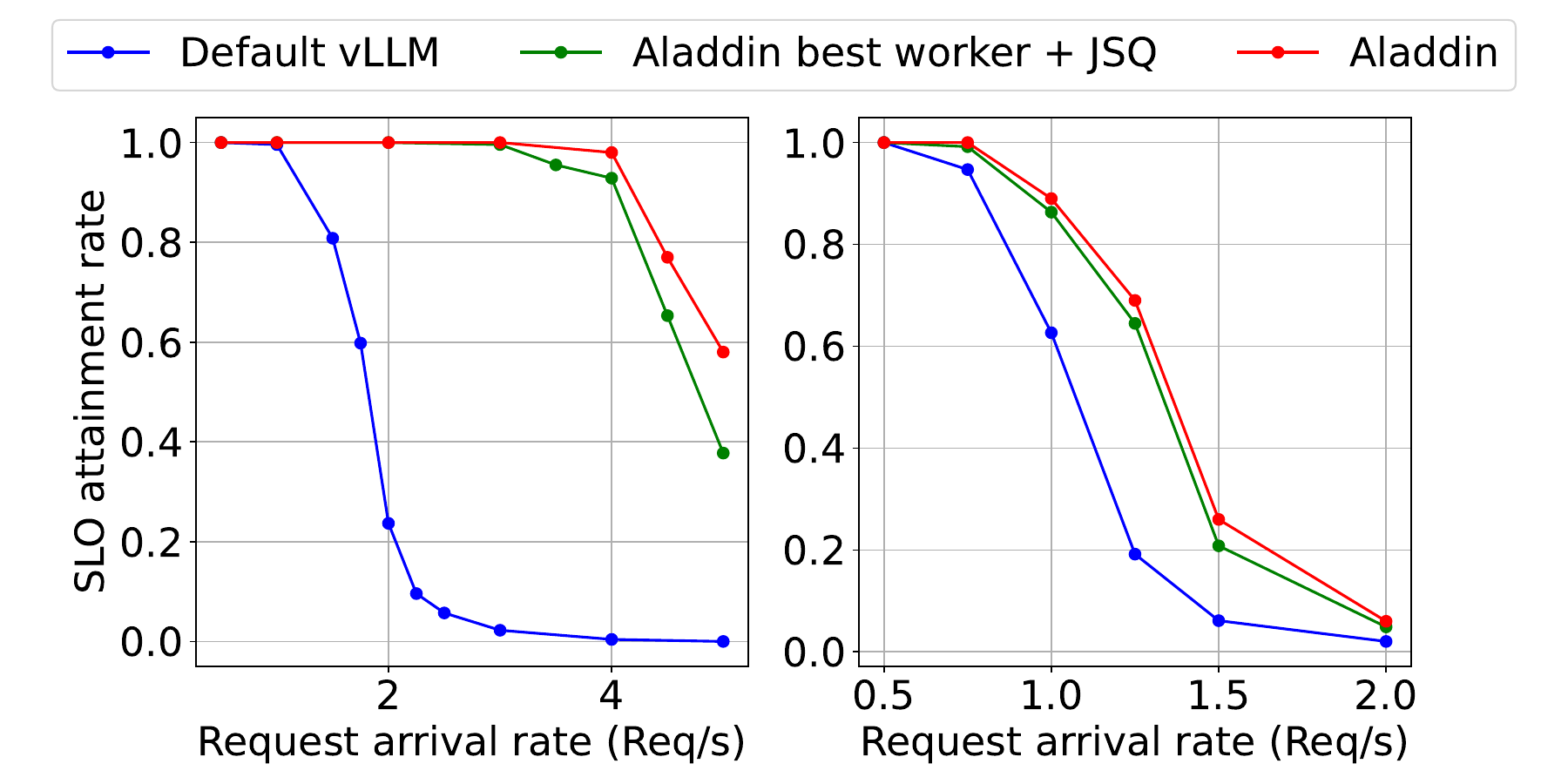}
        \caption{The end-to-end SLO attainment rate, (left): LlaMa2-13b, (right): LlaMa2-70b}
        \label{fig:slo_a100}
    \end{subfigure}
    \hfill
    \begin{subfigure}[b]{0.4\textwidth}
        \centering
        \includegraphics[width=\textwidth]{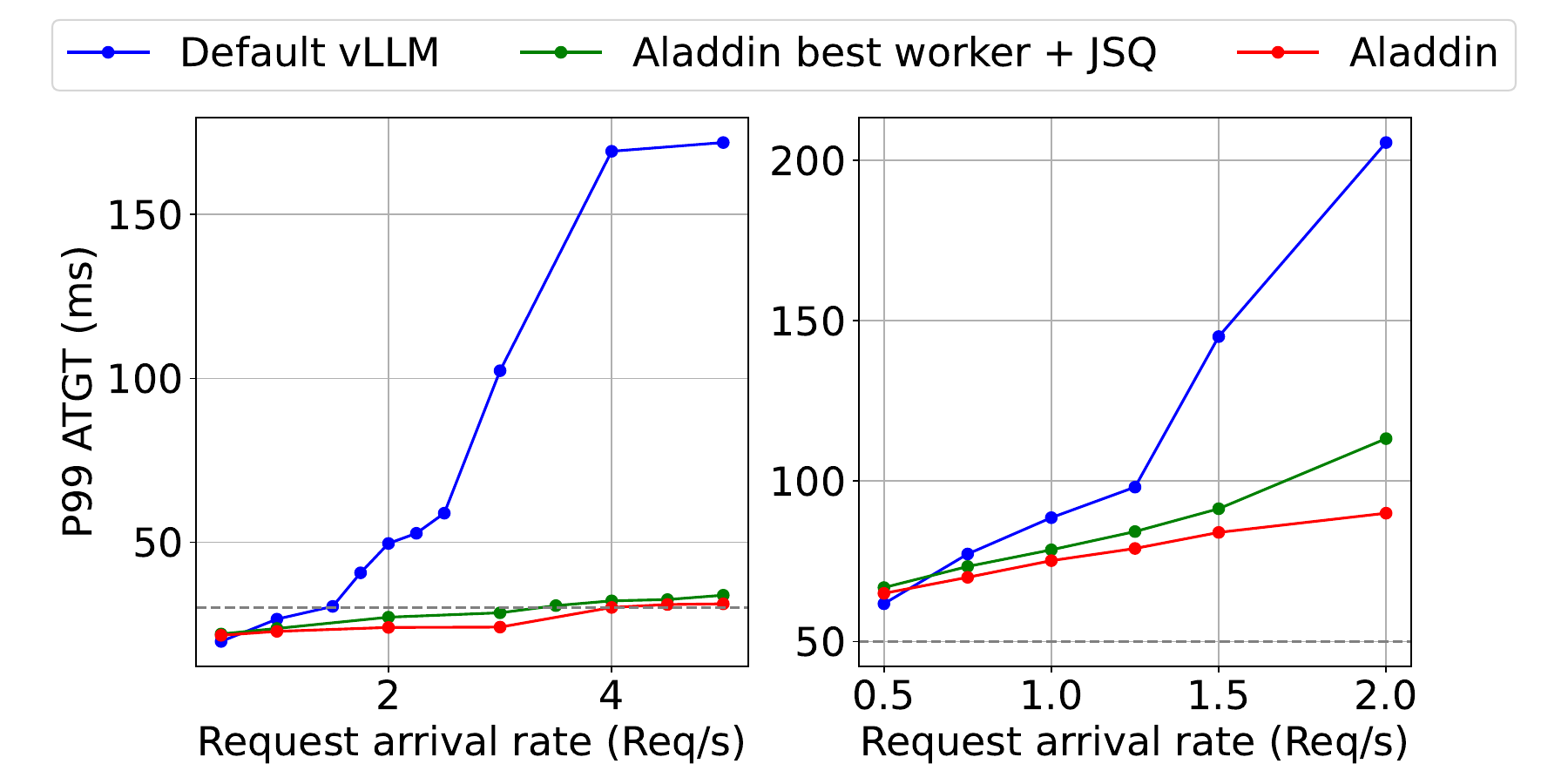}
        \caption{The end-to-end P99 ATGT, (left): LlaMa2-13b, (right): LlaMa2-70b}
        \label{fig:atgt_a100}
    \end{subfigure}
    \caption{End to end experiments on A100 testbed}
    \label{fig:endtoenda100}
\end{figure}

\subsection{Experimental Setup}
\textbf{Testbed setup.} We test the performance of Aladdin on high-end GPU servers with 4 A100 80GB GPUs connected with PCIe. Each machine has two  Intel Xeon Platinum 8380 processors and 512GB RAM.  
To validate the generalization of Aladdin from both a computation perspective and communication perspective, we also evaluate Aladdin on the GPU servers with 4 V100 32GB GPUs connected with NVLink. Each machine has two Intel Xeon Gold 6230 processors and 128GB RAM. We also do a large-scale simulation for the high-demand request arrival situation.\\
\textbf{Models and SLOs.} Our target is to prove Aladdin reduces the cost for transformer-based LLMs. To validate that Aladdin can accurately model the performance metrics of most models. We evaluate Aladdin on Llama2 series~\cite{touvron2023llama} models from 7B to 70B. The model, testbed information, and SLOs are shown in Table~\ref{tab:eval model}. Note that the prefill latency SLOs are the approximated inference latency for the model's context length (4096 tokens) for each testbed. The selection of decode latency SLO is according to the individual request inference latency. We guarantee that the batch inference latency of each request won't exceed the individual inference latency for $1.3$ times. \\
\textbf{Workload.} For the end-to-end performance evaluation in Section~\ref{sec:endtoend}, we first collect the prompts from users of ShareGPT\_V3\_unfiltered\_cleaned\_split dataset~\cite{sharegpt}, then submit the prompts follows a Poisson distribution.  The outputs are generated by each evaluated model with a temperature of $0$ and a maximum output token limit of $2048$. For the large-scale simulation in Section~\ref{sec:simlulation}, we use the same prompts' lengths as those collected from ShareGPT~\cite{sharegpt} in Section~\ref{sec:endtoend} as the prompt lengths. Then, we predict the output length based on the output length CDF of the responses generated in Section~\ref{sec:endtoend}'s end-to-end evaluations for each model.

\begin{table}[h]
\caption{The LLM information and testbed allocation}
  \small
  \centering
  \begin{tabular}{cccc}
    \toprule
    \multirow{2}* {\textbf{Model}}  &\multirow{2}* {\textbf{Testbed} } & \textbf{Prefill}&\textbf{Decode}    \\ 
    &  & \textbf{SLO(ms)} &  \textbf{SLO(ms)}  \\  
    \midrule
    Llama2-chat 70b & A100 & 1600 &75 \\
    Llama2-chat 13b  & A100, V100 & 600, 800 &30, 50 \\
    Llama2-chat 7b & A100, V100 & 400, 800 &15, 30  \\

  \bottomrule
  \end{tabular}
  \label{tab:eval model}
\end{table}

\begin{figure}
    \centering
    \begin{subfigure}[b]{0.4\textwidth}
        \centering
        \includegraphics[width=\textwidth]{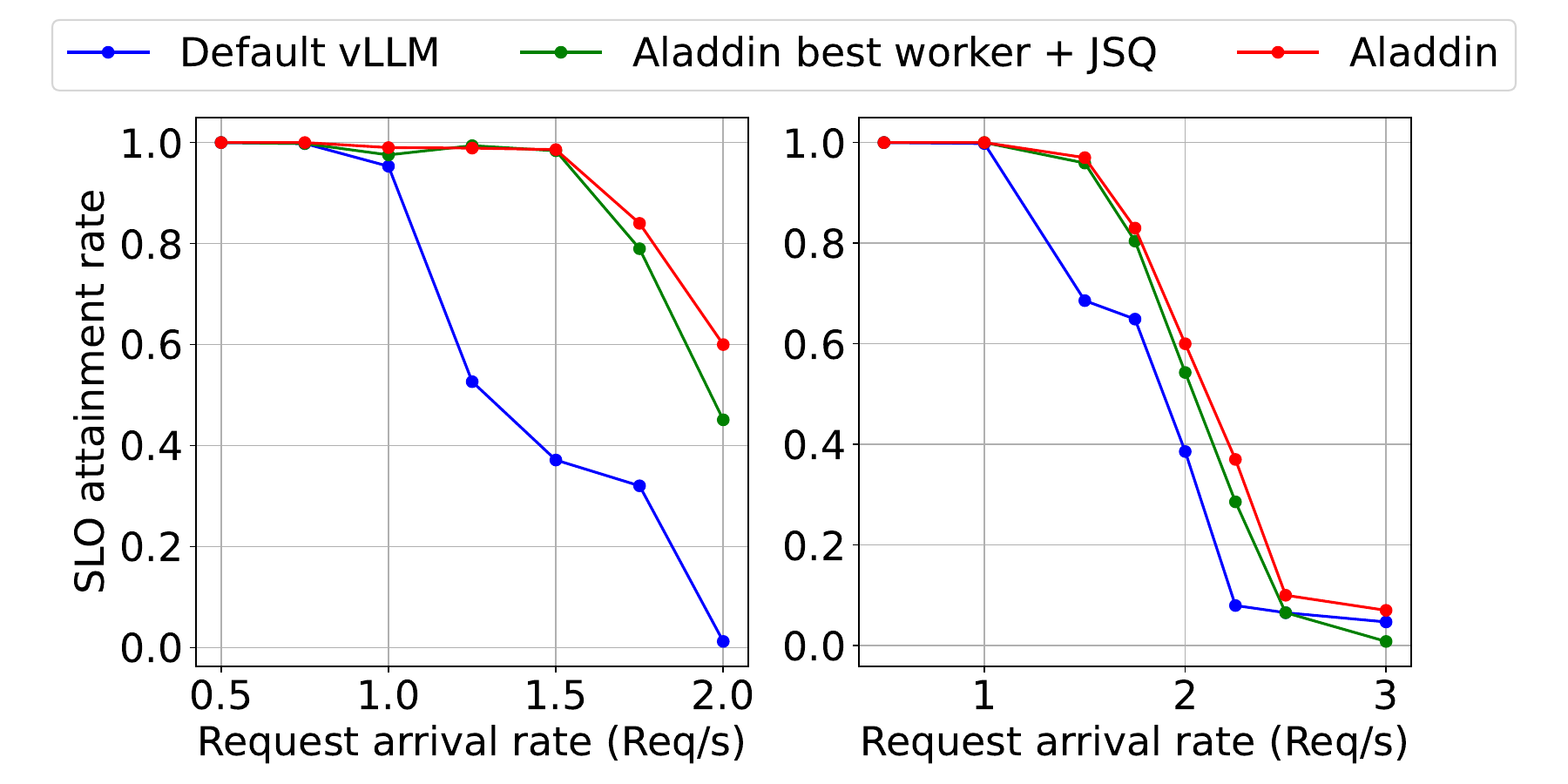}
        \caption{The end-to-end SLO attainment rate, (left): LlaMa2-7b, (right): LlaMa2-13b}
        \label{fig:slo_v100}
    \end{subfigure}
    \hfill
    \begin{subfigure}[b]{0.4\textwidth}
        \centering
        \includegraphics[width=\textwidth]{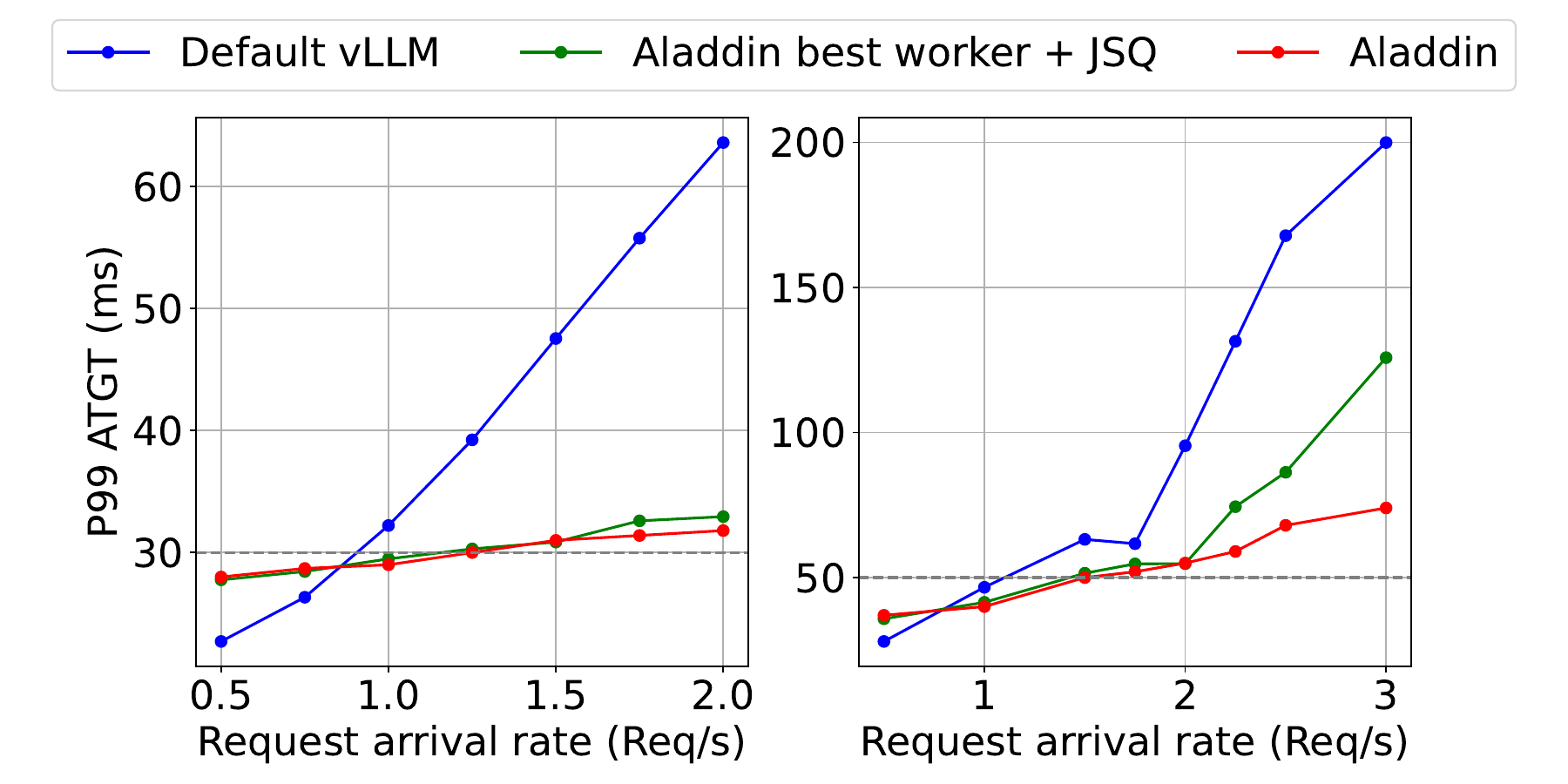}
        \caption{The end-to-end P99 ATGT, (left): LlaMa2-7b, (right): LlaMa2-13b}
        \label{fig:atgt_v100}
    \end{subfigure}
    \caption{End to end experiments on V100 testbed}
    \label{fig:endtoendv100}
\end{figure}

Because there is no available trace of LLM inference that includes the arrival time of each request, we simulate the request arrival stream using a Poisson distribution. We need to validate that Aladdin improves performance in both high-demand and low-demand scenarios. To evaluate the performance of Aladdin with varying demands, we tune the average arrival rate $\lambda$ to simulate different request demands. \\
\textbf{Metrics.} Since our final target is to reduce the cost of the inference service, we use the number of GPUs required to achieve a certain SLO attainment rate as the main metric. In Section~\ref{sec:simlulation}, we evaluate the total GPU number required with different request arrival rates. In Section~\ref{sec:endtoend}, As the total resources are limited for the real testbed evaluation, we evaluate the performance of Aladdin with the SLO attainment rate and the P99 ATGT in different request arrival rates. The SLO is attained when both TTFT and ATGT latency meet the requirement. \\ 
\textbf{Baselines.} Aladdin is a cluster-level scheduler. The performance improvement achieved by Aladdin is orthogonal to the improvements achieved by single-server optimization techniques such as split-phase inference~\cite{splitwise, distserve} or page attention~\cite{vllm}. These single-server optimization techniques use naive cluster scheduling like JSQ. Previous work~\cite{interference} adopts the power-of-two scheduling for request placement. However, it is suboptimal for request placement and cannot guarantee a high SLO attainment rate. We compared Aladdin's request placement with JSQ and power-of-two algorithms with different GPUs and different serving scenarios.

\begin{figure}
    \centering
    \includegraphics[width=1\linewidth]{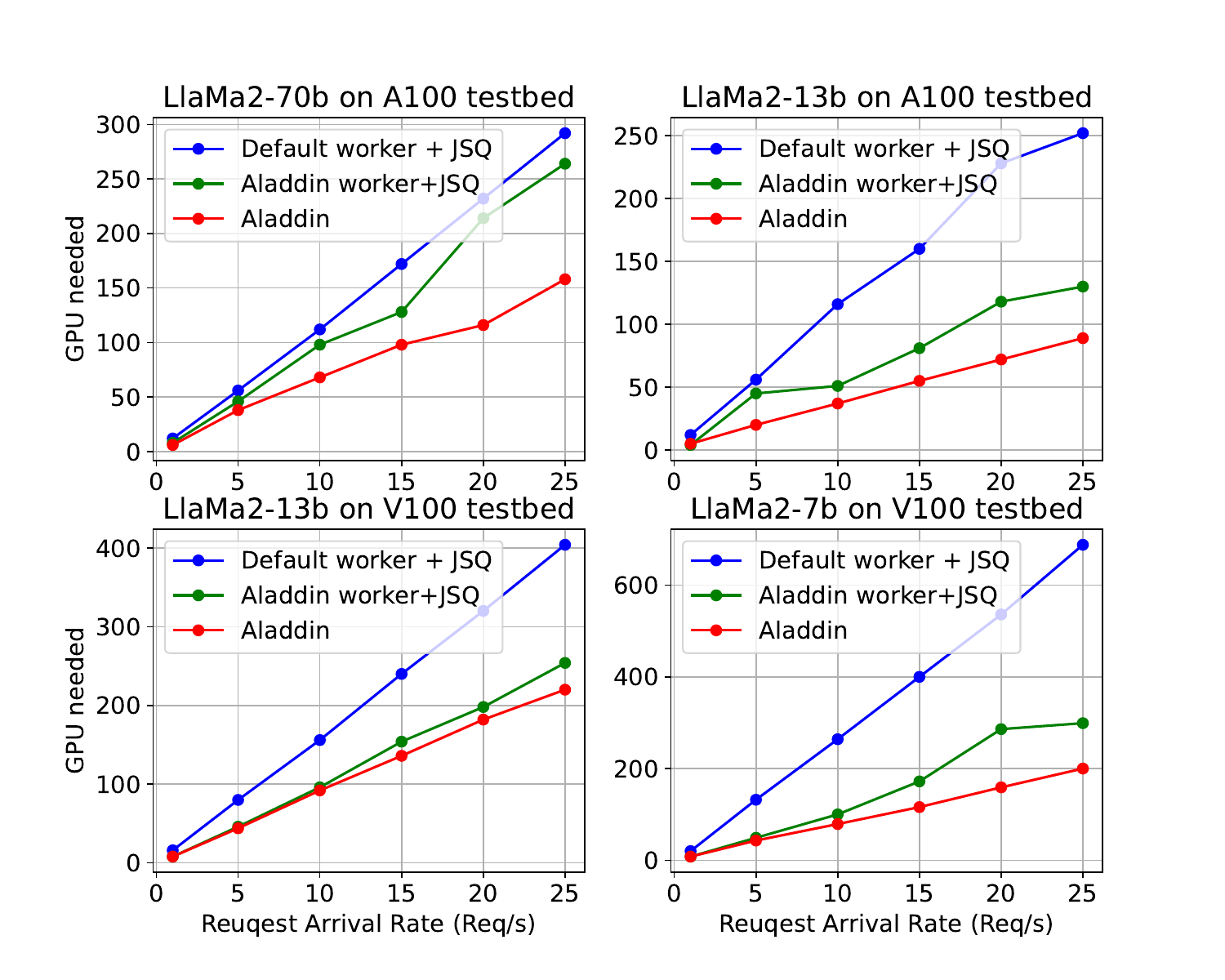}
    \caption{Simulation of the total GPU number needed with the mixed prefill and decode setting.}
    \label{fig:simuvllm}
\end{figure}

\subsection{Performance Model Validation}
\label{sec:model validate}
Request placement and worker configuration depend on accurate predictions of performance metrics. In this section, we evaluate the model's accuracy by comparing the predicted metrics to the measured actual metrics. 

In Section~\ref{sec:modeling}, we model the latency of prefill phase and decode phase separately because the two phases have different characteristics. In the evaluation, we evaluate the prefill and decode latency separately for different input and context lengths. In our prefill latency model, the prefill time of a batch size only corresponds to the total input length of all requests in the batch, not related to the batch size. In our experiment, we test different batch sizes $1, 2, 4, 8$ with the same total input length within a batch to validate this formulation.  We only evaluated the LlaMa2-70b model on the A100 testbed because our V100 testbed could not load the 70b model (around $140$GB) even with all GPUs ($32$GB*$4$). Figure~\ref{fig:prefill_latency_a100} and Figure~\ref{fig:prefill_latency_v100} shows the results on A100 and V100 testbeds. The maximum prefill latency prediction error is less than $4\%$. The shaded area is the prediction interval, which represents the estimation of the range in which future observations are likely to fall. Results indicate the maximum error of the prediction interval compared with our prediction is less than $10$ tokens.

In the decode latency model, the iteration time is linear with respect to both the batch size and the total context length within the batch, not related to the context length of each request in the batch. This means that regardless of whether the context length of each request in a batch is long or short, the decoding latency will be the same when the sum of the context lengths of all requests and the batch size is the same. In our experiment, for the same batch size, we use the same sum of context length but different context length distributions for all requests in a batch to validate this formulation. Results are presented in Figure~\ref{fig:decode limit a100} and Figure~\ref{fig:decode limit v100}. Similar to the prefill latency prediction, the prediction error is less than $5\%$. For the prediction interval, the error is less than $300$ tokens for all context tokens in the batch.

The KV cache usage to the context length is the most accurate metric in our performance models. According to Figure~\ref{fig:kv_usage}, the prediction error is less than $1\%$, and the prediction interval is just aligned with the prediction model. Note that the KV cache usage is not affected by the testbed; it is only related to the model. Generally speaking, the larger the model is, the more KV cache is needed for the same context length. However, from Figure~\ref{fig:kv_usage}, we can see that for the same context length, Llama2-13b requires more KV cache than Llama2-70b. This is because Llama2 7b and 13b adopt multi-head attention, while the 70b model adopts grouped-query attention~\cite{gqa}, which shares key and value pairs within each group.

\begin{figure}
    \centering
    \includegraphics[width=1\linewidth]{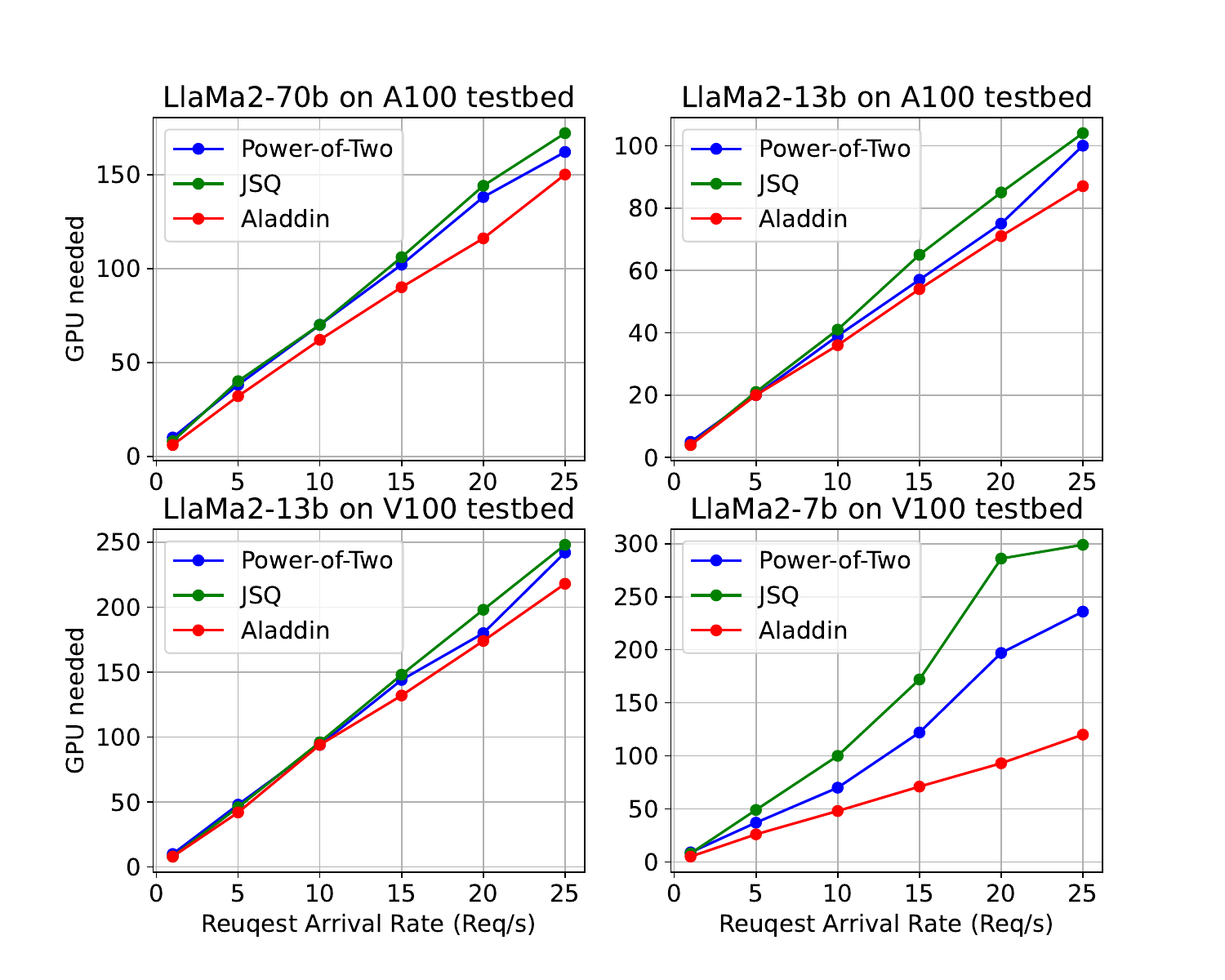}
    \caption{Simulation of the total GPU number needed for the decode phase of the split-phase inference setting}
    \label{fig:simusplit}
\end{figure}

\subsection{End-to-End Performance}
\label{sec:endtoend}
We evaluate Aladdin's end-to-end performance by comparing it with baselines on our A100 and V100 testbeds. In this experiment, requests arrived on Aladdin in a stream format following Poisson distribution. We use ShareGPT\cite{sharegpt} dataset for the conversation content. The baseline we select is the default vLLM, with all GPUs (4 GPUs) on each machine in one worker. Since the performance improvement achieved by Aladdin is gained both from request placement and optimal worker configuration, we configure vLLM with the optimal worker configuration and adopt JSQ for the request placement to do the ablation study. Table~\ref{tab:best configuration} reveals the best worker configuration for different models on different testbeds.

\begin{table}[h]
\caption{Optimal worker configuration for different models and different GPUs for ShareGPT dataset}
  \small
  \centering
  \begin{tabular}{cccc}
    \toprule
    \multirow{2}* {\textbf{Model}}   & \textbf{A100}&\textbf{V100}    \\ 
    & \textbf{(GPUs/worker)} & \textbf{(GPUs/worker)} &   \\  
    \midrule
    Llama2-70b-chat-hf & 2 & N/A  \\
    Llama2-13b-chat-hf  & 1 & 2 \\
    Llama2-7b-chat-hf & 1 & 1   \\

  \bottomrule
  \end{tabular}
  \label{tab:best configuration}
\end{table}

The results of A100 testbed are shown in Figure~\ref{fig:endtoenda100}. For the LlaMa2-70b model, Aladdin reduces the SLO violation rate by up to $3.5X$ compared with the default vLLM setting. Compared with the best worker configuration with JSQ placement, Aladdin only improved the SLO attainment rate by up to $19\%$. This is because there are totally two workers for the LlaMa2-70b model, which limits the improvement in the SLO attainment rate. However, Aladdin significantly reduces the P99 ATGT by up to $40\%$ compared with JSQ, as shown in Figure~\ref{fig:atgt_a100}'s right side. The results for the LlaMa2-13b model are distinct from the 70b model. The optimal worker configuration for the 13b on the A100 testbed is one GPU according to Table~\ref{tab:best configuration}. There are four workers in total for the request placement. So Aladdin improves the SLO attainment rate by up to $51\%$ compared with JSQ, but only has minor P99 ATGT improvement. The results of the V100 testbed are described in Figure~\ref{fig:endtoendv100}. The difference is when the request arrival rate is low, the P99 ATGT of baseline default vLLM output performs the performance with optimal worker configuration. This is because when the arrival rate is low, the batch effect is not significant, and the worker with more GPUs has higher computing power than the worker with fewer GPUs. Nevertheless, in those arrival rates, both baselines and Aladdin fulfill all requests SLOs. The higher ATGT won't further improve the SLO attainment rate.
Note that we don't include the P99 TTFT because vLLM~\cite{vllm} preempts the decode batch with the prefill batch when new requests arrive, making the ATGT more easily violate the SLO.

\subsection{Large-Scale Simulation}
\label{sec:simlulation}
We conducted a simulation for the high-demand request arrival scenario. In this simulation, we evaluated Aladdin's performance with split-phase inference and the default vLLM inference setting. To show the direct cost savings of Aladdin, we simulate the GPU number required for P100 SLO-guaranteed inference serving at the different request arrival rates. \\
\textbf{Default Continuous Batching Inference.} In Figure~\ref{fig:simuvllm}, we compared vLLM with baselines in Section~\ref{sec:endtoend}. Results indicate that Aladdin reduces the LLM serving cost by up to $71\%$ and $40\%$ compared with the default vLLM and JSQ with Aladdin optimal workers.\\
\textbf{Split-Phase Inference.} Previous work~\cite{splitwise, distserve, interference} split the prefill phase and decode phase into different instances. Split-phase serving maintains a group of prefill workers and a group of decode workers, as shown in Figure~\ref{fig:workflow_split}. According to the results of DistServe~\cite{distserve}, the majority of GPU resources are scheduled for the decode workers. Since the scheduling for prefill instances is trivial with known prompt lengths, we only simulate the GPU number required for the decode phase instance. The baselines are JSQ adopted by DistServe~\cite{distserve} and the Power-of-Two algorithm adopted by previous work~\cite{interference}. Results indicate that Aladdin reduces the GPU number required for the SLO-guaranteed decode phase by up to $60\%$ and $49\%$ compared with JSQ and Power-of-Two algorithm.

\subsection{Scheduling Overhead}
\label{sec:sche_overhead}
The scheduling overhead can be a problem in high-demand scenarios. For the scheduling latency, each scheduler's scheduling latency is predictable based on the request arrival rate since the time complexity of the best-fit bin packing algorithm is O(nlogn).
\begin{figure}
    \centering
    \includegraphics[width=0.6\linewidth]{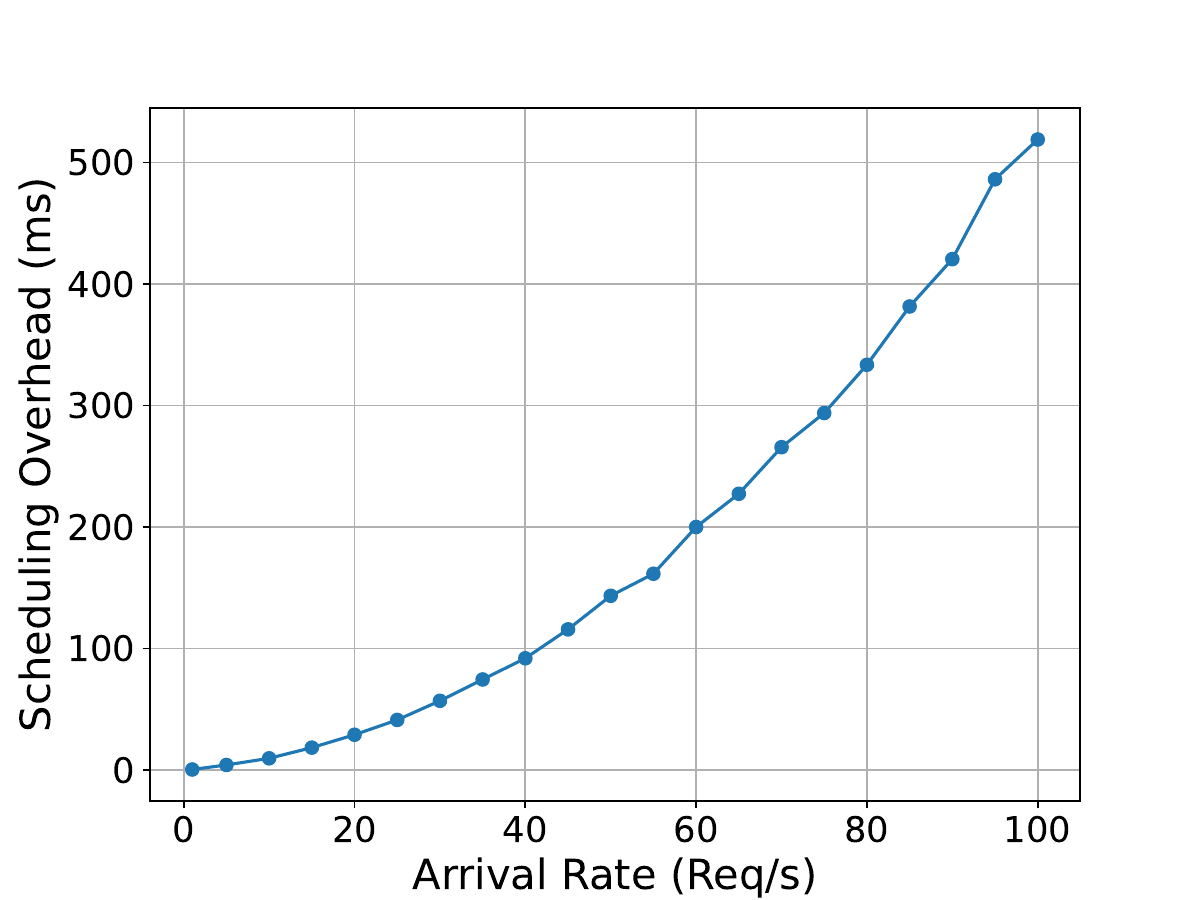}
    \caption{The bin packing algorithm running time with different arrival rates}
    \label{fig:overhead}
\end{figure}
Figure~\ref{fig:overhead} shows the scheduling overhead in centralized scheduling. According to the results, with a request arrival rate of around $25$ requests per second as we adopted in Section~\ref{sec:simlulation}. The scheduling overhead is less than $50$ ms, which is acceptable. However, if the arrival rate is very high or the scheduling latency limit is very strict, we can follow Appendix~\ref{sec:dist sched} to adopt the distributed grouped scheduling. 
\section{Related Work}
\textbf{LLM Inference Performance Modeling.} To improve the LLM inference metrics, the first step is to model the performance of the LLM inference process. Previous work~\cite{cheap_estimate} estimates the prefill and decode runtime based on the floating point operations during the inference. However, they focus on the inference for a single query. The performance metrics prediction of~\cite{cheap_estimate} is only based on the model, which lacks the consideration of the hardware adaption. DistServe~\cite{distserve} models the latency of prefill and decode phases for a batch of queries, also according to the model architecture, e.g., general matrix multiply operations. However, they model the inference latency mainly for the inference worker configuration instead of request placement. As far as we know, there is no existing research that studies the inference latency for a batch of requests with varying input and output lengths.   \\
\textbf{Inference Serving Quality.} The unique prefill-decode phase and the varying output length of autoregressive inference lead to the specialty of LLM inference SLOs. Because the first token generation time is only related to the input length, it makes the first token generation time predictable. There is a common understanding that the time to the first token SLO can be set as a constant~\cite{splitwise, distserve, interference, Sarathi-Serve}. However, most of the previous work~\cite{splitwise, interference, Sarathi-Serve} adopts the time between tokens as the SLO metric for the decode phase, which is overly strict and does not directly affect the user's experience, as discussed in Section~\ref{sec:slo}. Only considering the time between tokens SLO also harms the fairness of LLM inference~\cite{llmfairness}. The average token generation time SLO we use in this paper directly affects the quality of service and achieves better fairness for users.\\
\textbf{LLM Inference Systems.} Recent work on LLM serving systems can be categorized into three classes. The first category is application-level optimization~\cite{efficiently_scale}, where Continuous batching~\cite{orca} and page attention~\cite{vllm, strati2024dejavu} optimize the batch efficiency of LLM inference. Chunked-prefill~\cite{sarathi, Sarathi-Serve} balances the prefill and decode to improve LLM inference throughput. The second category is inference worker optimization~\cite{exegpt}; Splitwise~\cite{splitwise} adopts heterogeneous GPUs to handle different bottlenecks in prefill and decode phases. Previous work~\cite{distserve, interference} designed search algorithms to find the optimal worker configuration to achieve the best per-GPU goodput. The third class is about workload scheduling; recent work~\cite{andes, sche_out_pred} focuses on the scheduling of queries to improve the QoE or throughput. However, they lack consideration for resource management. 

\section{Conclusion}
We propose Aladdin, an adaptive LLM serving system that effectively scale and configures computing resources and optimally places inference queries to minimize serving costs while fulfilling SLOs. In this paper, we first deduce the performance models of the batched prefill and decode phases in LLM inference. Then, we predict the minimal computing resources required along with the corresponding worker configuration and request allocation. Results show that Aladdin reduced LLM serving costs by up to $71\%$ compared to state-of-the-art baselines.

{ \bibliographystyle{plain}
\bibliography{sample}}
\clearpage

\appendix
\section{Distributed Scheduling}
\label{sec:dist sched}

The scheduling time requirement of inference serving is in milliseconds. In a high-demand situation, the scheduling overhead is too large to place the requests in the target iteration for the centralized scheduler. We design a distributed scheduler for this case that harnesses the pattern of input and output length of requests.

With a high arrival rate, the worker required for inference service is linear to the request arrival rate as discussed in Section~\ref{sec: adapt_workload}. Hence we can randomly sample the arrived requests into groups using round robin. Then for each group, we only place the requests within the group to the workers corresponding to this group. While the arrival rate is $r_a$, if Group $i$ is corresponding to $N_i$ workers, the arrival rate of Group $i$ is $\frac{N_i}{N_w}r_a$. For this distributed scheduling mechanism, the scheduling demand of Group $i$ is $\frac{N_i}{N_w}$ of the total demand. Therefore, the scheduling latency is reduced. 

The selection of group numbers is non-trivial. With more request groups, the scheduling overhead is low because each group has fewer requests. However, more workers are needed to fulfill the SLOs because the workers' utilization is reduced compared with centralized scheduling. Assume half of the groups require one extra worker to fulfill the SLOs according to the probability that the requests placed to these groups require more computing resources, to limit the resource error to less than $e$ in percentage, each group should be equipped with at least $\frac{1}{2e}$ workers. For example, if we want the extra worker required for serving compared with centralized scheduling is $10\%$, then each scheduling group should have at least five workers. For the scheduling latency, each scheduler's scheduling latency is predictable based on the request arrival rate since the time complexity of the best-fit bin packing algorithm is O(nlogn). To guarantee both the scheduling latency and extra resource error, the request arrival rate of each scheduling group $r_i$ is constrained by:
\begin{equation}
    \label{eq:group_arrival_limit}
    \begin{split}
        &\frac{1}{2e}\leq r_i \leq r(T_s), \ i=1,2,\dots, N_{group},\\
        &\sum r_i = r_a, \ \ \ \ \ \ \ \ \ \  i=1,2,\dots, N_{group},
    \end{split}
\end{equation}
where $N_{group}$ is the group number, $T_s$ is the scheduling latency limit. The $r(t)$ is the function of the arrival rate limit to the scheduling latency. 

\begin{figure}[H]
    \centering
    \includegraphics[width=1\linewidth]{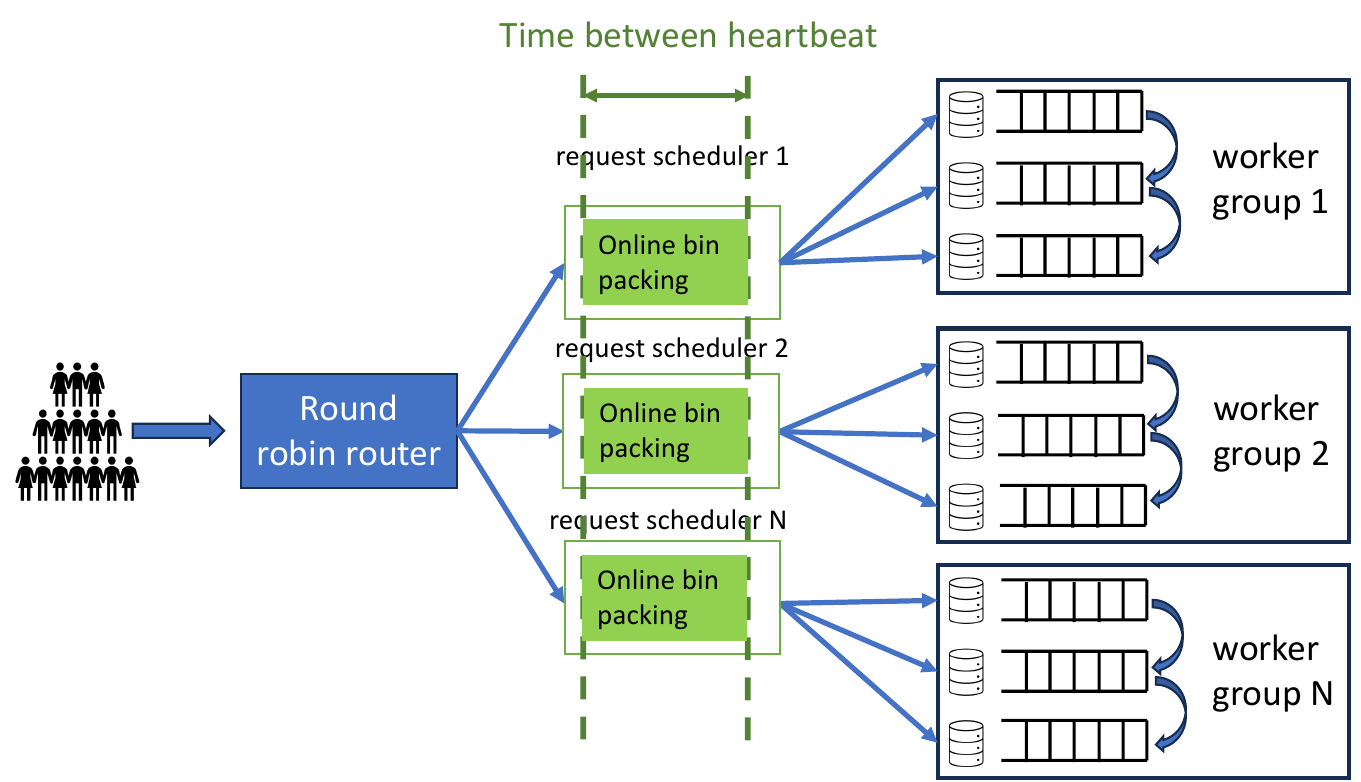}
    \caption{Distributed scheduling using request sampling.}
    \label{fig:dist_sche}
\end{figure}

With the distributed scheduling, the end-to-end co-adaptive scheduling of Aladdin is described in Figure~\ref{fig:dist_sche}. The first step is to predict the optimal worker configuration according to Section~\ref{sec:worker configuration} and the corresponding performance models based on Section~\ref{sec:modeling}. Given the request arrival rate, we predict the total worker number $N_w$ using Eq.~\ref{eq:total_worker_pre} and search for the group number $N_{group}$ using Eq.~\ref{eq:group_arrival_limit}. Then we use a round-robin router to route arriving requests to groups of schedulers. Finally, each scheduler packs requests to their corresponding workers using Algorithm~\ref{alg:heuristic}.

\end{document}